\documentclass[aps,prl,twocolumn,groupedaddress]{revtex4-1}
\usepackage{graphicx}
\usepackage{dcolumn}
\usepackage{bm}
\usepackage[english]{babel}
\usepackage{epsfig}
\usepackage[mathlines]{lineno}
\usepackage{amsmath,amssymb,graphicx,color}
\usepackage{multibib}

\begin{document}

\title{Enhanced emission extraction and selective readout of NV centers with all-dielectric nanoantennas}

\author{Alexander~E. Krasnok$^{1}$, Alex Maloshtan$^{2}$, Dmitry N. Chigrin$^{3}$, Yuri~S.
Kivshar$^{1,4}$, and Pavel~A. Belov$^{1}$}
\address{
$^{1}$~ITMO University, St. Petersburg 197101, Russia\\
$^{2}$~Institute of Physics of the National Academy of Science, Minsk, Belarusы\\
$^{3}$~Institute of Physics (1A), RWTH Aachen University, 52056 Aachen, Germany\\
$^{4}$~Nonlinear Physics Center, Research School of Physics and Engineering, Australian National University, Canberra ACT 0200, Australia}

\begin{abstract}
We propose a novel approach to facilitate a readout processes of isolated negatively charged nitrogen-vacancy (NV) centers based on the concept of all-dielectric nanoantennas. We reveal that all-dielectric nanoantenna can significantly enhance both the emission rate and emission extraction efficiency of a photoluminescence signal from a single NV center in a diamond nanoparticle placed on a dielectric substrate. We prove that the proposed approach provides high directivity, large Purcell factor, and efficient beam steering, thus allowing an efficient far-field initialization and readout of several NV centers separated by subwavelength distances.
\end{abstract}

\maketitle

\section*{Introduction}

Quantum emitters play an important role in many quantum-optics applications, ranging from
near-field microscopy~\cite{Novotny_Hecht_book} and sensing at the nanoscale~\cite{Neumann2013, Doherty2014, Lukin_08} to
single-photon emission devices~\cite{Kim13, Buckley12, Aharonovich2011,Weinfurter2000}, as well as from quantum information processing~\cite{Hartmann13, Hartmann2013, Hanson2013} and magnetic-resonance imaging with sub-nanometer
resolution~\cite{Yacoby_13, Yacoby_14} to biomedical applications~\cite{Schirhagl2013, Fann_07}. As a matter of fact, many realizations of quantum emitters have been demonstrated by now, including fluorescent molecules~\cite{Valeur_Book, Anger2006}, semiconductor quantum dots~\cite{Koch_12,Buckley12, Benson_2004}, and defect centers in solids~\cite{Aharonovich2011}. One of the most promising realizations of quantum emitters is based on negatively charged nitrogen-vacancy (NV) centers in diamond~\cite{Xu_12, Bakr_13,Wrachtrup_09}. Importantly, an electronic spin state of such a NV center can be readout by optical pulses~\cite{Jelezko2004a}, and it is also possible manipulating coherently spin states by applying microwave radiation~\cite{Wrachtrup2006b}. Additionally, the NV center demonstrates extremely long coherence time at room temperature, even under ambient conditions~\cite{Jacques_PRL_13, Wrachtrup_09}.

A typical scenario of an optical control of NV centers  includes optical initialization, microwave manipulation, and optical readout of electron spin states. The initialization is carried out by a green-laser pulse at 532~nm, which prepares the state $m_{\rm s}=0$, where $m_{\rm s}$ is the quantum number of the spin sub-levels. The microwave manipulation prepares an electronic spin state of a NV center in some superposition of the states $m_{\rm s}=0$ and $m_{\rm s}=\pm 1$. A single-shot readout of the resulting electronic spin state can be performed by two waves: by non-resonant green excitation (e.g., at 531 nm)~\cite{Manson2006} or by resonant excitation at the zero-phonon line (ZPL) at 637 nm~\cite{Wrachtrup2006b, Hanson11}. The photoluminescence spectrum of NV centers has an additional phononic sideband in the wavelength range of 600 nm to 780 nm and only up to 4$\%$ of photoluminescence goes into the ZPL emission at room temperature~\cite{Wolters2010}. Moreover, the radiation of a NV center placed on a dielectric substrate penetrates into the substrate. As a result, only a small part of the total radiation is collected by conventional confocal microscopy schemes~\cite{Wrachtrup_12}, so that it is highly desirable to increasing the emission extraction.

To enhance the emission rate, one can employ the Purcell effect~\cite{Maksymov_PRL_2013, 26}, known as a modification of
spontaneous decay rate $\gamma$ of a quantum emitter (in our case, a NV center) induced by the interaction with its
environment~\cite{Purcell_46, Maksymov_PRL_2013, Moerner_bowtie_09, Vahala_2003}. In the case of continuous pumping, the
Purcell effect leads to a change of the radiated power~\cite{58}. If the electromagnetic environment is lossless, the
Purcell factor describes a change in the total radiated power $P_{\rm rad}$ in the far-field zone:
\begin{equation}
\label{PurcellRad} F\equiv\frac{\gamma}{\gamma^0} = \frac{P_{\rm rad}}{P^{0}_{\rm rad}},
\end{equation}
where $P_{\rm rad}=\int_{\Omega}p(\theta,\varphi)d\Omega$ is the total power radiated to far field, $\theta$ and $\varphi$ are the angles of a spherical coordinate system, $d\Omega$ is the elementary solid angle, and the index 0 marks the
corresponding value in the free space. In the case of lossy environment, a part of the Purcell factor which is associated
with the radiation $F_{\rm rad}=\eta F$ can be defined. Here $\eta$ is a radiation efficiency of an emitter (a NV center in our case)~\cite{26, 58}:
\begin{equation}
\label{radeff}
\eta\equiv\frac{P_{\rm rad}}{P_{\rm rad}+P_{\rm loss}}=\frac{F_{\rm rad}}{F},
\end{equation}
where $P_{\rm loss}$ is a the power of dissipative losses in an inhomogeneous environment (cavity,
nanoantenna). The radiation enhancement at the ZPL frequency via the Purcell effect has been experimentally demonstrated for photonic-crystal cavities~\cite{Englund2010, Faraon2012}, micro-ring resonators~\cite{Fu2008, Barclay2011,
Beausoleil_11}, and metal substrates~\cite{Chi2011}.

The emission extraction efficiency can be improved using waveguiding or redirecting structures, like diamond
nanowire~\cite{Hausmann2010}, metallic apertures on a diamond surface~\cite{Choy2011} or plasmonic
nanoantenna~\cite{Bulu2011}. If NV center is located inside a diamond plate, emitted radiation can couple to guided modes
of the plate via the total internal reflection~\cite{Sage2012}.

A nanoantenna, for example plasmonic one, provides both emission rate and emission extraction efficiency enhancement.
Coupling of the emitter with a strongly localized near-field of the nanoantenna leads to increase of the Purcell factor
~\cite{Maksymov_PRL_2013, 26}. Moreover, the nanoantenna transforms a localized electromagnetic near-field into propagating
far-field with a desired radiation pattern~\cite{Novotny_10_NatPhot, Novotny_Hecht_book, 58}. The directivity of a
nanoantenna is defined as~\cite{26,58}:
\begin{equation}
\label{directivity}
D=\frac{4\pi}{P_{\rm rad}}\cdot{\rm
Max}[p(\theta,\varphi)],
\end{equation}
where ${\rm Max}[p(\theta,\varphi)]$ is the power flow density radiated in the direction of maximal flux density. Unfortunately, plasmonic nanoantennas usually possess strong dissipative losses resulting in low radiation efficiency ($\eta$) described by Eq.~(\ref{radeff})~\cite{Bozhevolnyi_NP_10,Tsakmakidis_NM_12, Bozhevolnyi_NP_14}.

Recently, it was suggested that all-dielectric nanoantennas can demonstrate much better performance in nanophotonics devices~\cite{58,7,BonodOE, BonodScRep} by employing the magnetic Mie resonances of high-index dielectric nanoparticles~\cite{Bozhevolnyi_SR_14, Kuznetsov, Chichkov_NL_12, Chichkov_PRB_10}. In particular, such all-dielectric nanoantennas demonstrate enhanced radiation efficiency in contrast to their plasmonic counterparts~\cite{58, 7, BonodOE, BonodScRep}. Moreover, all-dielectric nanoantennas may operate in the superdirective regime when higher-order multipole modes are excited, and the beam steering effect is observed~\cite{KrasnokNanoscale}.

In this paper, we study the emission extraction and readout of single NV centers and propose an alternative way to facilitate efficient readout processes by the simultaneous enhancement of the emission rate (via the Purcell
factor) and emission extraction efficiency at the ZPL wavelength. We employ the concept of all-dielectric optical nanoantennas as an effective tool for the NV center excitation and readout, and analyze numerically both emission rate and emission extraction efficiency of photoluminescence signal originating from a single NV center in a diamond placed on a
dielectric substrate close to the nanoantenna. Our approach also facilitates initialization of a single NV center in an
assemble with subwavelength separation among individual NV centers.

\begin{figure}[!t]
\centerline{\includegraphics[width=0.99\columnwidth]{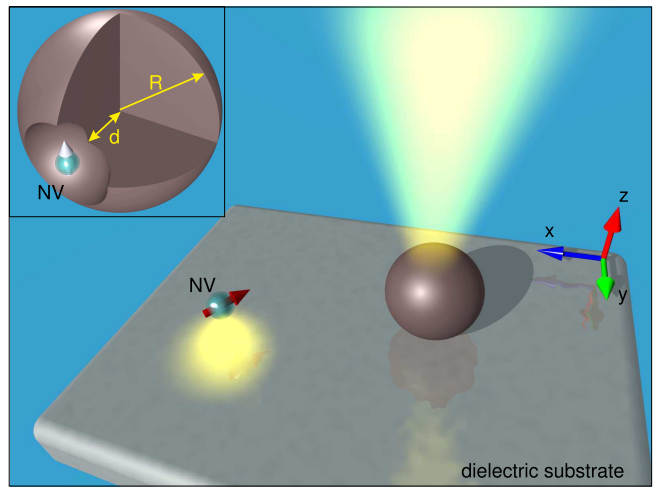}} \centering\caption{Schematic of an NV center emission {\em without} a nanoantenna (left; light is emitted into the substrate) and {\em with} a nanoantenna (right; light is directed by the antenna away from the substrate). Inset shows the geometry of the all-dielectric nanoantenna, a relative position of a diamond nanoparticle with a single NV center and dimensions.}
\label{fig:sketch}
\end{figure}


\section{1. Model}

The geometry under study is shown in Fig.\ref{fig:sketch}. As it was mentioned above the NV center primarily radiates into
substrate (see Fig.\ref{fig:sketch}, left). This decreases the power collected by confocal microscopy. A subwavelength
spherical nanoparticle of radius $R$ with a small notch is used as the all-dielectric nanoantenna. We consider a silicon
(Si) as a material of nanoparticle taking into account the real dissipative losses and frequency dispersion ~\cite{Palik}.
The notch is a result of cutting out of a sphere with radius $R_{\rm n}$ from the Si nanoparticle. The nanodiamond with NV center is located within the notch. The parameter $d$ is a distance between the nanoparticle center and the top edge of the notch. The nanoantenna and nanodiamond are both placed on a dielectric substrate with dielectric permittivity
$\varepsilon_{\rm s}$, as illustrated in Fig.\ref{fig:sketch}, right. The similar all-dielectric nanoantenna exited by a
point source located in the notch demonstrated high radiation efficiency and ability of beam steering at nanoscale, as
reported in Ref.~\cite{KrasnokNanoscale}. In this paper we use NV center as a source, detail the parameters of
all-dielectric nanoantenna and use the superdirectivity and beam steering effects for facilitating a far-field
initialization and readout of single nitrogen-vacancy center. It is important to mention, that proposed nanoantenna design
can be realized with available technologies. For example, similarly shaped nanoparticles were experimentally fabricated by
controlled deformation of a spherical shell~\cite{Sacanna2011, Yang08, Sacanna2011a}. Recent experimental progress in
spherical nanoparticles, nanoshells and semi-shells fabrication allows to expect realization of even more complex
nanoantennas of similar designs~\cite{VanDorpe2011, Halas2011, Giessen2012, Kuznetsov14, Chichkov_NC_14, Chichkov_Book,
Chichkov_APA}.

The shape and size of nanodiamond crystal as well as its environment affect the optical properties of NV
center~\cite{Beveratos2001,Greffet2011}. It has been shown that nominally identical nanocrystals can have high variance of
quantum efficiencies~\cite{KoenderinkNV13}. Moreover, it is widely known that a direct nuclear neighborhood of NV center
affects its properties~\cite{Smeltzer2011}. As a result, in our work we neglect these effects for simplicity and assume,
that a nanodiamond is preselected to contain single NV center with well defined
orientation~\cite{DiepLai2010,Geiselmann2013,Dolan2014}. An orientation of its dipole moments~\cite{Epstein2005} is assumed
to be perpendicular to the axis of the axial symmetry of the nanoantenna, as shown in Fig.\ref{fig:sketch} (see inset). We
assume that a nanodiamond is small enough to allow model the NV center as a point dipole emitter.
\begin{figure}[!t]
\centering \centerline{\includegraphics[width=0.85\columnwidth]{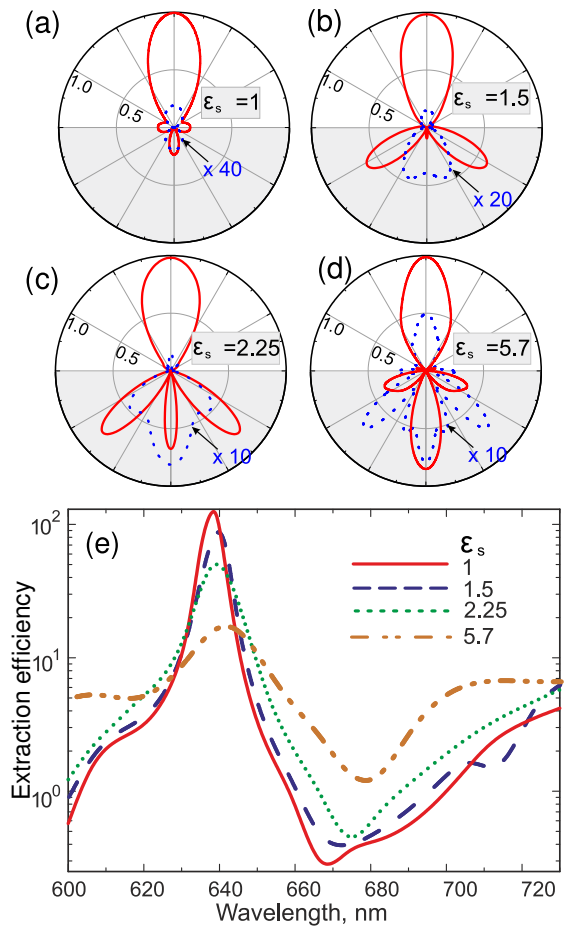}} \caption{The directivity patterns of NV
center radiation with (red solid curve) and without (blue dashed curve) all-dielectric nanoantenna at wavelength of 637 nm
(ZPL). The patterns are normalized to the maximum of NV center radiation with nanoantenna. The NV center is placed on 10 nm
above the dielectric substrate. The permittivity of substrate is $\varepsilon_{\rm s}$=1 (a), 1.5 (b), 2.25 glass (c), 5.7
diamond (d). (e) The power flow through 800 nm x 800 nm rectangular plane placed at distance 1000 nm above the substrate.
This dependence is normalized to the case of without nanoantenna.}
\label{fig:substrate}
\end{figure}

\section{2. Enhancement of emission extraction}

The proposed structure is quite complex for analytical study, especially taking into account the presence of the dielectric
substrate and permittivity dispersion. Therefore we have performed the full-wave numerical calculations using the CST
Microwave Studio. First, we have numerically optimized the nanoantenna geometry to maximize the radiation to the
superstrate at the frequency of ZPL ($\lambda=637$ nm). The found optimal parameters are equal to $R=R_{\rm n}=190$ nm and
$d=130$ nm. Then, we have analyzed the directivity (\ref{directivity}), radiation efficiency (\ref{radeff}) and Purcell
factor (\ref{PurcellRad}) of the optimized nanoantenna for substrates with different $\varepsilon_{\rm s}$. The directivity
patterns are presented in Fig.~\ref{fig:substrate}(a-d) (red solid curves). The scaled directivity patterns of NV center on
the substrate without nanoantenna are also presented in Fig.~\ref{fig:substrate}(a-d) (blue dashed curves). As follows from
these results, the nanoantenna provides high radiation directivity to superstrate direction even for a high dielectric
permittivity ($\varepsilon_{\rm s}=5.7$) of the substrate. Radiation into the substrate is significantly suppressed.

Further, we study the extraction efficiency, which we define as the ratio of the power flow into the superstrate with
nanoantenna and without it. We calculate the power flow through 800 nm x 800 nm rectangular plate which is placed at the
distance of 1000 nm over the substrate in the case of nanoantenna presence and normalized it to a case of without
nanoantenna. The results are presented in Fig.~\ref{fig:substrate}(e). For the case of substrate with $\varepsilon_{\rm
s}=1$ (air) we observed the increase of power flow in spectral region around the ZPL by two orders, versus the case of
nanoantenna absence. For the glass substrate ($\varepsilon_{\rm s}=2.25$) the value of the extraction efficiency is about
50 and even for the diamond substrate we observed the power flow increasing of 20 times.
\begin{figure}[!t] \centering
\centerline{\includegraphics[width=0.99\columnwidth]{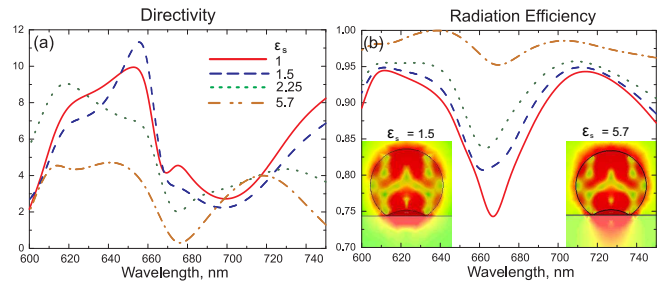}} \caption{(a) The directivity of radiation to the
straight up direction (in the positive direction of the axis $z$) for the different substrates. (b) The corresponding
radiation efficiency for the same. The insets depicts the electric field distributions at 637 nm with equal colorbar scale
for different substrates.}
\label{fig:efficiency}
\end{figure}

The origin of emission extraction enhancement is due to directivity and Purcell factor increasing, simultaneously. Lets
first consider the directivity of radiation to straight up direction and radiation efficiency for substrates with different
$\varepsilon_{\rm s}$. The results are presented in the Fig.\ref{fig:efficiency}(a) and (b), respectively. The directivity
of the all-dielectric nanoantenna in straight up direction strongly depends on $\varepsilon_{\rm s}$. For the case of
$\varepsilon_{\rm s}=1$ we obtain the directivity up to 10 at the spectral range around ZPL (see
Fig.\ref{fig:efficiency}(a)). For the glass ($\varepsilon_{\rm s}=2.25$) and diamond ($\varepsilon_{\rm s}=5.7$) substrates
we have obtained the straight up directivity at the frequency of ZPL close to 7 and 5, respectively. As it shown in the
Fig.\ref{fig:efficiency}(b) all-dielectric nanoantenna has the very high efficiency ($>75\%$) for the considered substrates
in all spectral range. This is due to the low dissipation losses of silicon at this frequencies. The radiation efficiency
increases with the increasing permittivity $\varepsilon_{\rm s}$ due to the electric field localization in the substrate
(see insets on the Fig.\ref{fig:efficiency}(b)).
\begin{figure}[!t] \centering
\centerline{\includegraphics[width=0.5\textwidth]{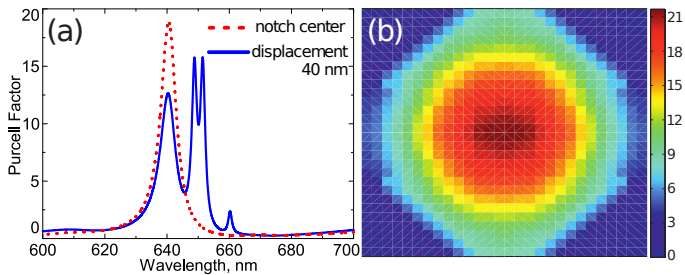}} \caption{(a) The Purcell factor for the NV center
placed 10 nm above the glass substrate ($\varepsilon_{\rm s}$=2.25) for symmetrical arrangement of NV center within the
notch (dashed line) and for shifted by 40 nm positions (solid blue line). (b) The Purcell factor map over plane parallel to
the substrate at ZPL wavelengths.} \label{fig:purcell} \end{figure}

The second mechanism which provides the high performance of the nanoantenna is the Purcell effect. The dependence of the
Purcell factor (for the glass substrate, $\varepsilon_{\rm s}=2.25$) for symmetrical arrangement of NV center within the
notch on the wavelength is shown in the Fig.~\ref{fig:purcell}(a) by red dashed curve. This dependence has the peak around
the ZPL (wavelength is 637 nm). The maximum of Purcell factor at the ZPL for the glass substrate ($\varepsilon_{\rm
s}=2.25$) and optimized geometrical parameters is about 20. This value decreases with increasing the substrate permittivity (not shown in the Fig.~\ref{fig:purcell}). In the cases of the air ($\varepsilon_{\rm s}=1$) and diamond ($\varepsilon_{\rm s}=5.7$) substrates the values of maximum of the Purcell factor are about 50 and 15, respectively. For non-symmetrical
arrangement of the NV center in the notch the additional peaks appeared as shown in the Fig.~\ref{fig:purcell}(a) by blue
solid curve. These peaks correspond to the higher multipole modes excitation, in full accordance to the results of
Ref.~\cite{KrasnokNanoscale}. The value of Purcell factor at the frequency of ZPL decreases with emitter shifting as shown
in the Fig.\ref{fig:purcell}(b). However, the enhancing effect through the Purcell factor is present within the entire
volume of the notch.
\begin{figure}[!b] \centering \centerline{\includegraphics[width=0.5\textwidth]{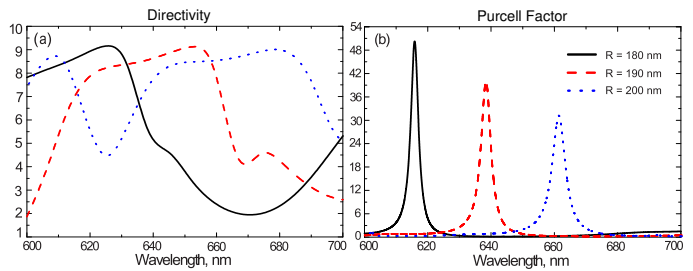}}
\caption{The directivity (a) and Purcell factor (b) dependence on the radius of nanoparticle ($R$) with fixed notch radius. The results correspond to the case of substrate absence ($\varepsilon_{\rm s}=1$).}
\label{fig:tuning}
\end{figure}

An important feature of the proposed all-dielectric nanoantenna is the ability to tune the operating frequency with respect
to geometrical parameters. The Fig.~\ref{fig:tuning} shows that the peaks of directivity and Purcell factor shift to the
longer wavelengths with increasing the nanoparticle size ($R$). The changing of the radius by 10 nm leads to a shift of
operating wavelength by 25 nm. Thus the range of possible applications of this all-dielectric nanoantenna is not limited to
NV centers.

\section{3. Selective readout of single NV centers}

A challenging problem in the quantum information processing is selective excitation or readout in a system of two or more
single photon emitters, separated by a distance smaller than the diffraction limit~\cite{Englund_NL_13, Xiangping_2013}.
Here we show that our approach can overcome this limitation and provide the effective control over the several NV centers
in the notch. In Ref.~\cite{KrasnokNanoscale} the effect of far field radiation profile steering which is caused by the
emitter offset inside the notch was investigated. It is interesting to observe the opposite situation when the several NV
centers are irradiated by one external plane wave.
\begin{figure}[!t] \centering
\centerline{\includegraphics[width=0.85\columnwidth]{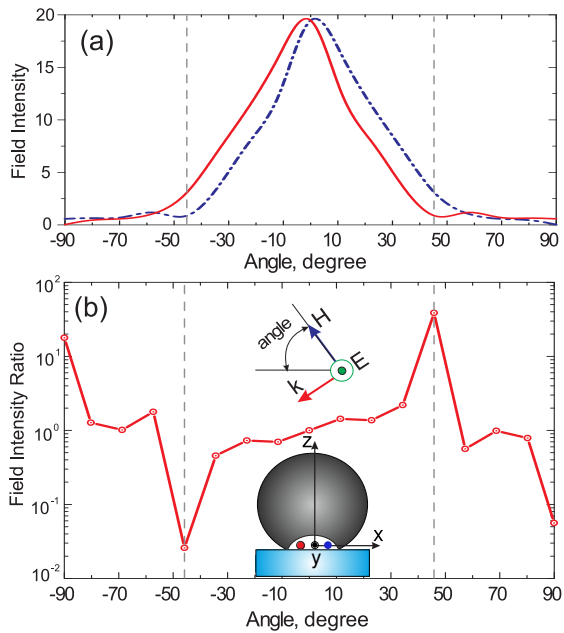}} \caption{\label{fig:excitation}(a) Dependence of
electric field intensity ($|E_{y}|^2$) at NV center positions on the angle of wave incidence. The distance between NV centers
is 80 nm. The dipole moments of the NV centers are oriented perpendicular to the separation direction. The polarization of
incident plane wave ($\lambda=637$ nm) is parallel to the emitters dipole moments (along the axis $y$). (b) The ratio of
the electric field intensities at two NV centers. The maximum values of ratio are reached at angle $\simeq\pm45$ degrees of
incidence.} \end{figure}

We consider the red resonance pulses (wavelength is 637 nm) which are usually used for NV center electron spin state single
shot readout~\cite{Manson2006}. We assume that two nanodiamonds with single NV centers are placed on the top of a silica
substrate ($\varepsilon_{\rm s}=2.25)$ inside the notch of the all-dielectric nanoantenna. The distance between NV centers
is 80 nm. The dipole moments of NV centers are oriented perpendicular to the separation direction. The polarization of
incident electric field is parallel to the emitters dipole moments (along the axis $y$, see inset in the
Fig.\ref{fig:excitation}). The dependence of electric field intensity ($|E_{y}|^2$) at NV center positions on the angle of
incidence is shown on the Fig.~\ref{fig:excitation}(a). The ratio of these intensities is presented in the
Fig.~\ref{fig:excitation}(b). The maximum of ratio is about $~20$ and is reached at the angles of incidence
$\simeq\pm45^\circ$. It is known that in the dipole approximation, the probability of photon absorption by dipole emitter
is proportional to the intensity of local electric field~\cite{Epstein2005}. Thus, we conclude that if the both NV centers
polarizabilities are the same, the excitation probability of first NV center may be 20 times greater than the excitation
probability of the second one. The same analysis can be performed for 531 nm excitation wavelength. We believe that
proposed design can be useful for systems with several quantum emitters for selective spin state far-field initialization
and readout.


\section*{Conclusions}

We have demonstrated numerically that all-dielectric optical nanoantenna can enhance significantly the intensity of a photoluminescence signal coming from a single NV center facilitating readout of its electron spin states. Our design results in (i) redistribution of the emitted radiation towards superstrate with high directivity and (ii) emission rate enhancement due to a high value of the Purcell factor. The spectral characteristic of the nanoantenna can be adjusted in a wide range by an appropriate choice of geometrical parameters. We have shown that the proposed approach can be used to simplify an initialization of a single NV center in an assemble with a subwavelength separation among individual NV centers by utilizing the efficient beam steering. We believe that the proposed design can be useful for different systems with several quantum emitters for selective spin-state far-field initialization and readout.

\section*{Acknowledgments}

The authors acknowledge useful discussions with P. Melentiev and I.V. Iorsh, and they also grateful to D.S. Filonov and P. V. Kapitanova for their interest in this work. This work was supported by the Ministry of Education and Science of the Russian Federation (GOSZADANIE 2014/190, Zadanie No. 3.561.2014/K, 14.584.21.0009 10), the Government of the Russian Federation (grant 074-U01), the Russian Foundation for Basic Research, the Dynasty Foundation (Russia), and the Australian Research Council.


\begin{thebibliography}{78}%
\makeatletter
\providecommand \@ifxundefined [1]{%
 \@ifx{#1\undefined}
}%
\providecommand \@ifnum [1]{%
 \ifnum #1\expandafter \@firstoftwo
 \else \expandafter \@secondoftwo
 \fi
}%
\providecommand \@ifx [1]{%
 \ifx #1\expandafter \@firstoftwo
 \else \expandafter \@secondoftwo
 \fi
}%
\providecommand \natexlab [1]{#1}%
\providecommand \enquote  [1]{``#1''}%
\providecommand \bibnamefont  [1]{#1}%
\providecommand \bibfnamefont [1]{#1}%
\providecommand \citenamefont [1]{#1}%
\providecommand \href@noop [0]{\@secondoftwo}%
\providecommand \href [0]{\begingroup \@sanitize@url \@href}%
\providecommand \@href[1]{\@@startlink{#1}\@@href}%
\providecommand \@@href[1]{\endgroup#1\@@endlink}%
\providecommand \@sanitize@url [0]{\catcode `\\12\catcode `\$12\catcode
  `\&12\catcode `\#12\catcode `\^12\catcode `\_12\catcode `\%12\relax}%
\providecommand \@@startlink[1]{}%
\providecommand \@@endlink[0]{}%
\providecommand \url  [0]{\begingroup\@sanitize@url \@url }%
\providecommand \@url [1]{\endgroup\@href {#1}{\urlprefix }}%
\providecommand \urlprefix  [0]{URL }%
\providecommand \Eprint [0]{\href }%
\providecommand \doibase [0]{http://dx.doi.org/}%
\providecommand \selectlanguage [0]{\@gobble}%
\providecommand \bibinfo  [0]{\@secondoftwo}%
\providecommand \bibfield  [0]{\@secondoftwo}%
\providecommand \translation [1]{[#1]}%
\providecommand \BibitemOpen [0]{}%
\providecommand \bibitemStop [0]{}%
\providecommand \bibitemNoStop [0]{.\EOS\space}%
\providecommand \EOS [0]{\spacefactor3000\relax}%
\providecommand \BibitemShut  [1]{\csname bibitem#1\endcsname}%
\let\auto@bib@innerbib\@empty
\bibitem [{\citenamefont {Novotny}\ and\ \citenamefont
  {Hecht}(2006)}]{Novotny_Hecht_book}%
  \BibitemOpen
  \bibfield  {author} {\bibinfo {author} {\bibfnamefont {L.}~\bibnamefont
  {Novotny}}\ and\ \bibinfo {author} {\bibfnamefont {B.}~\bibnamefont
  {Hecht}},\ }\href@noop {} {\emph {\bibinfo {title} {Principles of
  Nano-Optics}}}\ (\bibinfo  {publisher} {Cambridge University Press},\
  \bibinfo {year} {2006})\BibitemShut {NoStop}%
\bibitem [{\citenamefont {Neumann}\ \emph {et~al.}(2013)\citenamefont
  {Neumann}, \citenamefont {Jakobi}, \citenamefont {Dolde}, \citenamefont
  {Burk}, \citenamefont {Reuter}, \citenamefont {Waldherr}, \citenamefont
  {Honert}, \citenamefont {Wolf}, \citenamefont {Brunner}, \citenamefont
  {Shim}, \citenamefont {Suter}, \citenamefont {Isoya},\ and\ \citenamefont
  {Wrachtrup}}]{Neumann2013}%
  \BibitemOpen
  \bibfield  {author} {\bibinfo {author} {\bibfnamefont {P.}~\bibnamefont
  {Neumann}}, \bibinfo {author} {\bibfnamefont {I.}~\bibnamefont {Jakobi}},
  \bibinfo {author} {\bibfnamefont {F.}~\bibnamefont {Dolde}}, \bibinfo
  {author} {\bibfnamefont {C.}~\bibnamefont {Burk}}, \bibinfo {author}
  {\bibfnamefont {R.}~\bibnamefont {Reuter}}, \bibinfo {author} {\bibfnamefont
  {G.}~\bibnamefont {Waldherr}}, \bibinfo {author} {\bibfnamefont
  {J.}~\bibnamefont {Honert}}, \bibinfo {author} {\bibfnamefont
  {T.}~\bibnamefont {Wolf}}, \bibinfo {author} {\bibfnamefont {A.}~\bibnamefont
  {Brunner}}, \bibinfo {author} {\bibfnamefont {J.~H.}\ \bibnamefont {Shim}},
  \bibinfo {author} {\bibfnamefont {D.}~\bibnamefont {Suter}}, \bibinfo
  {author} {\bibfnamefont {S.~J.}\ \bibnamefont {Isoya}}, \ and\ \bibinfo
  {author} {\bibfnamefont {J.}~\bibnamefont {Wrachtrup}},\ }\href@noop {}
  {\bibfield  {journal} {\bibinfo  {journal} {Nano Letters}\ }\textbf {\bibinfo
  {volume} {13}},\ \bibinfo {pages} {2738} (\bibinfo {year}
  {2013})}\BibitemShut {NoStop}%
\bibitem [{\citenamefont {Doherty}\ \emph {et~al.}(2014)\citenamefont
  {Doherty}, \citenamefont {Struzhkin}, \citenamefont {Simpson}, \citenamefont
  {McGuinness}, \citenamefont {Meng}, \citenamefont {Stacey}, \citenamefont
  {Karle}, \citenamefont {Hemley}, \citenamefont {Manson}, \citenamefont
  {Hollenberg},\ and\ \citenamefont {Prawer}}]{Doherty2014}%
  \BibitemOpen
  \bibfield  {author} {\bibinfo {author} {\bibfnamefont {M.~W.}\ \bibnamefont
  {Doherty}}, \bibinfo {author} {\bibfnamefont {V.~V.}\ \bibnamefont
  {Struzhkin}}, \bibinfo {author} {\bibfnamefont {D.~A.}\ \bibnamefont
  {Simpson}}, \bibinfo {author} {\bibfnamefont {L.~P.}\ \bibnamefont
  {McGuinness}}, \bibinfo {author} {\bibfnamefont {Y.}~\bibnamefont {Meng}},
  \bibinfo {author} {\bibfnamefont {A.}~\bibnamefont {Stacey}}, \bibinfo
  {author} {\bibfnamefont {T.~J.}\ \bibnamefont {Karle}}, \bibinfo {author}
  {\bibfnamefont {R.~J.}\ \bibnamefont {Hemley}}, \bibinfo {author}
  {\bibfnamefont {N.~B.}\ \bibnamefont {Manson}}, \bibinfo {author}
  {\bibfnamefont {L.~C.}\ \bibnamefont {Hollenberg}}, \ and\ \bibinfo {author}
  {\bibfnamefont {S.}~\bibnamefont {Prawer}},\ }\href@noop {} {\bibfield
  {journal} {\bibinfo  {journal} {Physical Review Letters}\ }\textbf {\bibinfo
  {volume} {112}},\ \bibinfo {pages} {047601} (\bibinfo {year}
  {2014})}\BibitemShut {NoStop}%
\bibitem [{\citenamefont {Taylor}\ \emph {et~al.}(2008)\citenamefont {Taylor},
  \citenamefont {Cappellaro}, \citenamefont {Childress}, \citenamefont {Jiang},
  \citenamefont {Budker}, \citenamefont {Hemmer}, \citenamefont {Yacoby},
  \citenamefont {Walsworth},\ and\ \citenamefont {Lukin}}]{Lukin_08}%
  \BibitemOpen
  \bibfield  {author} {\bibinfo {author} {\bibfnamefont {J.~M.}\ \bibnamefont
  {Taylor}}, \bibinfo {author} {\bibfnamefont {P.}~\bibnamefont {Cappellaro}},
  \bibinfo {author} {\bibfnamefont {L.}~\bibnamefont {Childress}}, \bibinfo
  {author} {\bibfnamefont {L.}~\bibnamefont {Jiang}}, \bibinfo {author}
  {\bibfnamefont {D.}~\bibnamefont {Budker}}, \bibinfo {author} {\bibfnamefont
  {P.~R.}\ \bibnamefont {Hemmer}}, \bibinfo {author} {\bibfnamefont
  {A.}~\bibnamefont {Yacoby}}, \bibinfo {author} {\bibfnamefont
  {R.}~\bibnamefont {Walsworth}}, \ and\ \bibinfo {author} {\bibfnamefont
  {M.~D.}\ \bibnamefont {Lukin}},\ }\href@noop {} {\bibfield  {journal}
  {\bibinfo  {journal} {Nature Physics}\ }\textbf {\bibinfo {volume} {4}},\
  \bibinfo {pages} {810 } (\bibinfo {year} {2008})}\BibitemShut {NoStop}%
\bibitem [{\citenamefont {Kim}\ \emph {et~al.}(2013)\citenamefont {Kim},
  \citenamefont {Ko}, \citenamefont {Gong}, \citenamefont {Ko},\ and\
  \citenamefont {Cho}}]{Kim13}%
  \BibitemOpen
  \bibfield  {author} {\bibinfo {author} {\bibfnamefont {J.-H.}\ \bibnamefont
  {Kim}}, \bibinfo {author} {\bibfnamefont {Y.-H.}\ \bibnamefont {Ko}},
  \bibinfo {author} {\bibfnamefont {S.-H.}\ \bibnamefont {Gong}}, \bibinfo
  {author} {\bibfnamefont {S.-M.}\ \bibnamefont {Ko}}, \ and\ \bibinfo {author}
  {\bibfnamefont {Y.-H.}\ \bibnamefont {Cho}},\ }\href@noop {} {\bibfield
  {journal} {\bibinfo  {journal} {Scientific Reports}\ }\textbf {\bibinfo
  {volume} {3}},\ \bibinfo {pages} {2150} (\bibinfo {year} {2013})}\BibitemShut
  {NoStop}%
\bibitem [{\citenamefont {Buckley}\ \emph {et~al.}(2012)\citenamefont
  {Buckley}, \citenamefont {Rivoire},\ and\ \citenamefont
  {Vuckovic}}]{Buckley12}%
  \BibitemOpen
  \bibfield  {author} {\bibinfo {author} {\bibfnamefont {S.}~\bibnamefont
  {Buckley}}, \bibinfo {author} {\bibfnamefont {K.}~\bibnamefont {Rivoire}}, \
  and\ \bibinfo {author} {\bibfnamefont {J.}~\bibnamefont {Vuckovic}},\
  }\href@noop {} {\bibfield  {journal} {\bibinfo  {journal} {Rep. Prog. Phys.}\
  }\textbf {\bibinfo {volume} {75}},\ \bibinfo {pages} {126503} (\bibinfo
  {year} {2012})}\BibitemShut {NoStop}%
\bibitem [{\citenamefont {Aharonovich}\ \emph {et~al.}(2011)\citenamefont
  {Aharonovich}, \citenamefont {Castelletto}, \citenamefont {Simpson},
  \citenamefont {Su}, \citenamefont {Greentree},\ and\ \citenamefont
  {Prawer}}]{Aharonovich2011}%
  \BibitemOpen
  \bibfield  {author} {\bibinfo {author} {\bibfnamefont {I.}~\bibnamefont
  {Aharonovich}}, \bibinfo {author} {\bibfnamefont {S.}~\bibnamefont
  {Castelletto}}, \bibinfo {author} {\bibfnamefont {D.~A.}\ \bibnamefont
  {Simpson}}, \bibinfo {author} {\bibfnamefont {C.-H.}\ \bibnamefont {Su}},
  \bibinfo {author} {\bibfnamefont {A.~D.}\ \bibnamefont {Greentree}}, \ and\
  \bibinfo {author} {\bibfnamefont {S.}~\bibnamefont {Prawer}},\ }\href@noop {}
  {\bibfield  {journal} {\bibinfo  {journal} {Rep. Prog. Phys.}\ }\textbf
  {\bibinfo {volume} {74}},\ \bibinfo {pages} {076501} (\bibinfo {year}
  {2011})}\BibitemShut {NoStop}%
\bibitem [{\citenamefont {Kurtsiefer}\ \emph {et~al.}(2000)\citenamefont
  {Kurtsiefer}, \citenamefont {Mayer}, \citenamefont {Zarda},\ and\
  \citenamefont {Weinfurter}}]{Weinfurter2000}%
  \BibitemOpen
  \bibfield  {author} {\bibinfo {author} {\bibfnamefont {C.}~\bibnamefont
  {Kurtsiefer}}, \bibinfo {author} {\bibfnamefont {S.}~\bibnamefont {Mayer}},
  \bibinfo {author} {\bibfnamefont {P.}~\bibnamefont {Zarda}}, \ and\ \bibinfo
  {author} {\bibfnamefont {H.}~\bibnamefont {Weinfurter}},\ }\href@noop {}
  {\bibfield  {journal} {\bibinfo  {journal} {Phys. Rev. Lett.}\ }\textbf
  {\bibinfo {volume} {85}},\ \bibinfo {pages} {290} (\bibinfo {year}
  {2000})}\BibitemShut {NoStop}%
\bibitem [{\citenamefont {Rips}\ and\ \citenamefont
  {Hartmann}(2013)}]{Hartmann13}%
  \BibitemOpen
  \bibfield  {author} {\bibinfo {author} {\bibfnamefont {S.}~\bibnamefont
  {Rips}}\ and\ \bibinfo {author} {\bibfnamefont {M.~J.}\ \bibnamefont
  {Hartmann}},\ }\href@noop {} {\bibfield  {journal} {\bibinfo  {journal}
  {Phys. Rev. Lett.}\ }\textbf {\bibinfo {volume} {110}},\ \bibinfo {pages}
  {120503} (\bibinfo {year} {2013})}\BibitemShut {NoStop}%
\bibitem [{\citenamefont {Neumeier}\ \emph {et~al.}(2013)\citenamefont
  {Neumeier}, \citenamefont {Leib},\ and\ \citenamefont
  {Hartmann}}]{Hartmann2013}%
  \BibitemOpen
  \bibfield  {author} {\bibinfo {author} {\bibfnamefont {L.}~\bibnamefont
  {Neumeier}}, \bibinfo {author} {\bibfnamefont {M.}~\bibnamefont {Leib}}, \
  and\ \bibinfo {author} {\bibfnamefont {M.~J.}\ \bibnamefont {Hartmann}},\
  }\href@noop {} {\bibfield  {journal} {\bibinfo  {journal} {Phys. Rev. Lett.}\
  }\textbf {\bibinfo {volume} {111}},\ \bibinfo {pages} {63601} (\bibinfo
  {year} {2013})}\BibitemShut {NoStop}%
\bibitem [{\citenamefont {Childress}\ and\ \citenamefont
  {Hanson}(2013)}]{Hanson2013}%
  \BibitemOpen
  \bibfield  {author} {\bibinfo {author} {\bibfnamefont {L.}~\bibnamefont
  {Childress}}\ and\ \bibinfo {author} {\bibfnamefont {R.}~\bibnamefont
  {Hanson}},\ }\href@noop {} {\bibfield  {journal} {\bibinfo  {journal}
  {Nitrogen-vacancy centers: Physics and applications}\ }\textbf {\bibinfo
  {volume} {38}},\ \bibinfo {pages} {134} (\bibinfo {year} {2013})}\BibitemShut
  {NoStop}%
\bibitem [{\citenamefont {Grinolds}\ \emph {et~al.}(2013)\citenamefont
  {Grinolds}, \citenamefont {Hong}, \citenamefont {Maletinsky}, \citenamefont
  {Luan}, \citenamefont {Lukin}, \citenamefont {Walsworth},\ and\ \citenamefont
  {Yacoby}}]{Yacoby_13}%
  \BibitemOpen
  \bibfield  {author} {\bibinfo {author} {\bibfnamefont {M.~S.}\ \bibnamefont
  {Grinolds}}, \bibinfo {author} {\bibfnamefont {S.}~\bibnamefont {Hong}},
  \bibinfo {author} {\bibfnamefont {P.}~\bibnamefont {Maletinsky}}, \bibinfo
  {author} {\bibfnamefont {L.}~\bibnamefont {Luan}}, \bibinfo {author}
  {\bibfnamefont {M.~D.}\ \bibnamefont {Lukin}}, \bibinfo {author}
  {\bibfnamefont {R.~L.}\ \bibnamefont {Walsworth}}, \ and\ \bibinfo {author}
  {\bibfnamefont {A.}~\bibnamefont {Yacoby}},\ }\href@noop {} {\bibfield
  {journal} {\bibinfo  {journal} {Nature Physics}\ }\textbf {\bibinfo {volume}
  {9}},\ \bibinfo {pages} {215} (\bibinfo {year} {2013})}\BibitemShut {NoStop}%
\bibitem [{\citenamefont {Grinolds}\ \emph {et~al.}(2014)\citenamefont
  {Grinolds}, \citenamefont {Warner}, \citenamefont {Greve}, \citenamefont
  {Dovzhenko}, \citenamefont {Thiel}, \citenamefont {Walsworth}, \citenamefont
  {Hong}, \citenamefont {Maletinsky},\ and\ \citenamefont
  {Yacoby}}]{Yacoby_14}%
  \BibitemOpen
  \bibfield  {author} {\bibinfo {author} {\bibfnamefont {M.~S.}\ \bibnamefont
  {Grinolds}}, \bibinfo {author} {\bibfnamefont {M.}~\bibnamefont {Warner}},
  \bibinfo {author} {\bibfnamefont {K.~D.}\ \bibnamefont {Greve}}, \bibinfo
  {author} {\bibfnamefont {Y.}~\bibnamefont {Dovzhenko}}, \bibinfo {author}
  {\bibfnamefont {L.}~\bibnamefont {Thiel}}, \bibinfo {author} {\bibfnamefont
  {R.~L.}\ \bibnamefont {Walsworth}}, \bibinfo {author} {\bibfnamefont
  {S.}~\bibnamefont {Hong}}, \bibinfo {author} {\bibfnamefont {P.}~\bibnamefont
  {Maletinsky}}, \ and\ \bibinfo {author} {\bibfnamefont {A.}~\bibnamefont
  {Yacoby}},\ }\href@noop {} {\bibfield  {journal} {\bibinfo  {journal} {Nature
  Nanotechnology}\ }\textbf {\bibinfo {volume} {9}},\ \bibinfo {pages} {279}
  (\bibinfo {year} {2014})}\BibitemShut {NoStop}%
\bibitem [{\citenamefont {Schirhagl}\ \emph {et~al.}(2014)\citenamefont
  {Schirhagl}, \citenamefont {Chang}, \citenamefont {Loretz},\ and\
  \citenamefont {Degen}}]{Schirhagl2013}%
  \BibitemOpen
  \bibfield  {author} {\bibinfo {author} {\bibfnamefont {R.}~\bibnamefont
  {Schirhagl}}, \bibinfo {author} {\bibfnamefont {K.}~\bibnamefont {Chang}},
  \bibinfo {author} {\bibfnamefont {M.}~\bibnamefont {Loretz}}, \ and\ \bibinfo
  {author} {\bibfnamefont {C.~L.}\ \bibnamefont {Degen}},\ }\href@noop {}
  {\bibfield  {journal} {\bibinfo  {journal} {Annu Rev Phys Chem.}\ }\textbf
  {\bibinfo {volume} {65}},\ \bibinfo {pages} {83} (\bibinfo {year}
  {2014})}\BibitemShut {NoStop}%
\bibitem [{\citenamefont {Fu}\ \emph {et~al.}(2007)\citenamefont {Fu},
  \citenamefont {Lee}, \citenamefont {Chen}, \citenamefont {Lim}, \citenamefont
  {Wu}, \citenamefont {Lin}, \citenamefont {Wei}, \citenamefont {Tsao},
  \citenamefont {Chang},\ and\ \citenamefont {Fann}}]{Fann_07}%
  \BibitemOpen
  \bibfield  {author} {\bibinfo {author} {\bibfnamefont {C.-C.}\ \bibnamefont
  {Fu}}, \bibinfo {author} {\bibfnamefont {H.-Y.}\ \bibnamefont {Lee}},
  \bibinfo {author} {\bibfnamefont {K.}~\bibnamefont {Chen}}, \bibinfo {author}
  {\bibfnamefont {T.-S.}\ \bibnamefont {Lim}}, \bibinfo {author} {\bibfnamefont
  {H.-Y.}\ \bibnamefont {Wu}}, \bibinfo {author} {\bibfnamefont {P.-K.}\
  \bibnamefont {Lin}}, \bibinfo {author} {\bibfnamefont {P.-K.}\ \bibnamefont
  {Wei}}, \bibinfo {author} {\bibfnamefont {P.-H.}\ \bibnamefont {Tsao}},
  \bibinfo {author} {\bibfnamefont {H.-C.}\ \bibnamefont {Chang}}, \ and\
  \bibinfo {author} {\bibfnamefont {W.}~\bibnamefont {Fann}},\ }\href@noop {}
  {\bibfield  {journal} {\bibinfo  {journal} {PNAS}\ }\textbf {\bibinfo
  {volume} {104}},\ \bibinfo {pages} {727} (\bibinfo {year}
  {2007})}\BibitemShut {NoStop}%
\bibitem [{\citenamefont {Valeur}(2001)}]{Valeur_Book}%
  \BibitemOpen
  \bibfield  {author} {\bibinfo {author} {\bibfnamefont {B.}~\bibnamefont
  {Valeur}},\ }\href@noop {} {\emph {\bibinfo {title} {Molecular Fluorescence.
  Principles and Applications}}}\ (\bibinfo  {publisher} {Wiley-VCH},\ \bibinfo
  {year} {2001})\BibitemShut {NoStop}%
\bibitem [{\citenamefont {Anger}\ \emph {et~al.}(2006)\citenamefont {Anger},
  \citenamefont {Bharadwaj},\ and\ \citenamefont {Novotny}}]{Anger2006}%
  \BibitemOpen
  \bibfield  {author} {\bibinfo {author} {\bibfnamefont {P.}~\bibnamefont
  {Anger}}, \bibinfo {author} {\bibfnamefont {P.}~\bibnamefont {Bharadwaj}}, \
  and\ \bibinfo {author} {\bibfnamefont {L.}~\bibnamefont {Novotny}},\
  }\href@noop {} {\bibfield  {journal} {\bibinfo  {journal} {Phys. Rev. Lett.}\
  }\textbf {\bibinfo {volume} {96}},\ \bibinfo {pages} {113002} (\bibinfo
  {year} {2006})}\BibitemShut {NoStop}%
\bibitem [{\citenamefont {Kira}\ and\ \citenamefont {Koch}(2012)}]{Koch_12}%
  \BibitemOpen
  \bibfield  {author} {\bibinfo {author} {\bibfnamefont {M.}~\bibnamefont
  {Kira}}\ and\ \bibinfo {author} {\bibfnamefont {S.~W.}\ \bibnamefont
  {Koch}},\ }\href@noop {} {\emph {\bibinfo {title} {Semiconductor Quantum
  Optics}}}\ (\bibinfo  {publisher} {Cambridge University Press},\ \bibinfo
  {year} {2012})\BibitemShut {NoStop}%
\bibitem [{\citenamefont {Zwiller}\ \emph {et~al.}(2004)\citenamefont
  {Zwiller}, \citenamefont {Aichele},\ and\ \citenamefont
  {Benson}}]{Benson_2004}%
  \BibitemOpen
  \bibfield  {author} {\bibinfo {author} {\bibfnamefont {V.}~\bibnamefont
  {Zwiller}}, \bibinfo {author} {\bibfnamefont {T.}~\bibnamefont {Aichele}}, \
  and\ \bibinfo {author} {\bibfnamefont {O.}~\bibnamefont {Benson}},\
  }\href@noop {} {\bibfield  {journal} {\bibinfo  {journal} {New Journal of
  Physics}\ }\textbf {\bibinfo {volume} {96}},\ \bibinfo {pages} {76177}
  (\bibinfo {year} {2004})}\BibitemShut {NoStop}%
\bibitem [{\citenamefont {Hsua}\ and\ \citenamefont {Xu}(2012)}]{Xu_12}%
  \BibitemOpen
  \bibfield  {author} {\bibinfo {author} {\bibfnamefont {C.-H.}\ \bibnamefont
  {Hsua}}\ and\ \bibinfo {author} {\bibfnamefont {J.}~\bibnamefont {Xu}},\
  }\href@noop {} {\bibfield  {journal} {\bibinfo  {journal} {Nanoscale}\
  }\textbf {\bibinfo {volume} {4}},\ \bibinfo {pages} {5293} (\bibinfo {year}
  {2012})}\BibitemShut {NoStop}%
\bibitem [{\citenamefont {Mahfouz}\ \emph {et~al.}(2013)\citenamefont
  {Mahfouz}, \citenamefont {Floyd}, \citenamefont {Peng}, \citenamefont {Choy},
  \citenamefont {Loncarb},\ and\ \citenamefont {Bakr}}]{Bakr_13}%
  \BibitemOpen
  \bibfield  {author} {\bibinfo {author} {\bibfnamefont {R.}~\bibnamefont
  {Mahfouz}}, \bibinfo {author} {\bibfnamefont {D.~L.}\ \bibnamefont {Floyd}},
  \bibinfo {author} {\bibfnamefont {W.}~\bibnamefont {Peng}}, \bibinfo {author}
  {\bibfnamefont {J.~T.}\ \bibnamefont {Choy}}, \bibinfo {author}
  {\bibfnamefont {M.}~\bibnamefont {Loncarb}}, \ and\ \bibinfo {author}
  {\bibfnamefont {O.~M.}\ \bibnamefont {Bakr}},\ }\href@noop {} {\bibfield
  {journal} {\bibinfo  {journal} {Nanoscale}\ }\textbf {\bibinfo {volume}
  {5}},\ \bibinfo {pages} {11776} (\bibinfo {year} {2013})}\BibitemShut
  {NoStop}%
\bibitem [{\citenamefont {Balasubramanian}\ \emph {et~al.}(2009)\citenamefont
  {Balasubramanian}, \citenamefont {Neumann}, \citenamefont {Twitchen},
  \citenamefont {Markham}, \citenamefont {Kolesov}, \citenamefont {Mizuochi},
  \citenamefont {Isoya}, \citenamefont {Achard}, \citenamefont {Beck},
  \citenamefont {Tissler}, \citenamefont {Jacques}, \citenamefont {Hemmer},
  \citenamefont {Jelezko},\ and\ \citenamefont {Wrachtrup}}]{Wrachtrup_09}%
  \BibitemOpen
  \bibfield  {author} {\bibinfo {author} {\bibfnamefont {G.}~\bibnamefont
  {Balasubramanian}}, \bibinfo {author} {\bibfnamefont {P.}~\bibnamefont
  {Neumann}}, \bibinfo {author} {\bibfnamefont {D.}~\bibnamefont {Twitchen}},
  \bibinfo {author} {\bibfnamefont {M.}~\bibnamefont {Markham}}, \bibinfo
  {author} {\bibfnamefont {R.}~\bibnamefont {Kolesov}}, \bibinfo {author}
  {\bibfnamefont {N.}~\bibnamefont {Mizuochi}}, \bibinfo {author}
  {\bibfnamefont {J.}~\bibnamefont {Isoya}}, \bibinfo {author} {\bibfnamefont
  {J.}~\bibnamefont {Achard}}, \bibinfo {author} {\bibfnamefont
  {J.}~\bibnamefont {Beck}}, \bibinfo {author} {\bibfnamefont {J.}~\bibnamefont
  {Tissler}}, \bibinfo {author} {\bibfnamefont {V.}~\bibnamefont {Jacques}},
  \bibinfo {author} {\bibfnamefont {P.~R.}\ \bibnamefont {Hemmer}}, \bibinfo
  {author} {\bibfnamefont {F.}~\bibnamefont {Jelezko}}, \ and\ \bibinfo
  {author} {\bibfnamefont {J.}~\bibnamefont {Wrachtrup}},\ }\href@noop {}
  {\bibfield  {journal} {\bibinfo  {journal} {Nature Materials}\ }\textbf
  {\bibinfo {volume} {8}},\ \bibinfo {pages} {383 } (\bibinfo {year}
  {2009})}\BibitemShut {NoStop}%
\bibitem [{\citenamefont {Jelezko}\ \emph {et~al.}(2004)\citenamefont
  {Jelezko}, \citenamefont {Gaebel}, \citenamefont {Popa}, \citenamefont
  {Gruber},\ and\ \citenamefont {Wrachtrup}}]{Jelezko2004a}%
  \BibitemOpen
  \bibfield  {author} {\bibinfo {author} {\bibfnamefont {F.}~\bibnamefont
  {Jelezko}}, \bibinfo {author} {\bibfnamefont {T.}~\bibnamefont {Gaebel}},
  \bibinfo {author} {\bibfnamefont {I.}~\bibnamefont {Popa}}, \bibinfo {author}
  {\bibfnamefont {A.}~\bibnamefont {Gruber}}, \ and\ \bibinfo {author}
  {\bibfnamefont {J.}~\bibnamefont {Wrachtrup}},\ }\href {\doibase
  10.1103/PhysRevLett.92.076401} {\bibfield  {journal} {\bibinfo  {journal}
  {Phys. Rev. Lett.}\ }\textbf {\bibinfo {volume} {92}},\ \bibinfo {pages}
  {76401} (\bibinfo {year} {2004})}\BibitemShut {NoStop}%
\bibitem [{\citenamefont {Wrachtrup}\ and\ \citenamefont
  {Jelezko}(2006)}]{Wrachtrup2006b}%
  \BibitemOpen
  \bibfield  {author} {\bibinfo {author} {\bibfnamefont {J.}~\bibnamefont
  {Wrachtrup}}\ and\ \bibinfo {author} {\bibfnamefont {F.}~\bibnamefont
  {Jelezko}},\ }\href {\doibase 10.1088/0953-8984/18/21/S08} {\bibfield
  {journal} {\bibinfo  {journal} {Journal of Physics: Condensed Matter}\
  }\textbf {\bibinfo {volume} {18}},\ \bibinfo {pages} {S807} (\bibinfo {year}
  {2006})}\BibitemShut {NoStop}%
\bibitem [{\citenamefont {Dreau}\ \emph {et~al.}(2013)\citenamefont {Dreau},
  \citenamefont {Spinicelli}, \citenamefont {Maze}, \citenamefont {Roch},\ and\
  \citenamefont {Jacques}}]{Jacques_PRL_13}%
  \BibitemOpen
  \bibfield  {author} {\bibinfo {author} {\bibfnamefont {A.}~\bibnamefont
  {Dreau}}, \bibinfo {author} {\bibfnamefont {P.}~\bibnamefont {Spinicelli}},
  \bibinfo {author} {\bibfnamefont {J.~R.}\ \bibnamefont {Maze}}, \bibinfo
  {author} {\bibfnamefont {J.-F.}\ \bibnamefont {Roch}}, \ and\ \bibinfo
  {author} {\bibfnamefont {V.}~\bibnamefont {Jacques}},\ }\href@noop {}
  {\bibfield  {journal} {\bibinfo  {journal} {Phys. Rev. Lett.}\ }\textbf
  {\bibinfo {volume} {110}},\ \bibinfo {pages} {060502} (\bibinfo {year}
  {2013})}\BibitemShut {NoStop}%
\bibitem [{\citenamefont {Manson}\ \emph {et~al.}(2006)\citenamefont {Manson},
  \citenamefont {Harrison},\ and\ \citenamefont {Sellars}}]{Manson2006}%
  \BibitemOpen
  \bibfield  {author} {\bibinfo {author} {\bibfnamefont {N.}~\bibnamefont
  {Manson}}, \bibinfo {author} {\bibfnamefont {J.}~\bibnamefont {Harrison}}, \
  and\ \bibinfo {author} {\bibfnamefont {M.}~\bibnamefont {Sellars}},\
  }\href@noop {} {\bibfield  {journal} {\bibinfo  {journal} {Physical Review
  B}\ }\textbf {\bibinfo {volume} {74}},\ \bibinfo {pages} {104303} (\bibinfo
  {year} {2006})}\BibitemShut {NoStop}%
\bibitem [{\citenamefont {Robledo}\ \emph {et~al.}(2011)\citenamefont
  {Robledo}, \citenamefont {Childress}, \citenamefont {Bernien}, \citenamefont
  {Hensen}, \citenamefont {Alkemade},\ and\ \citenamefont {Hanson}}]{Hanson11}%
  \BibitemOpen
  \bibfield  {author} {\bibinfo {author} {\bibfnamefont {L.}~\bibnamefont
  {Robledo}}, \bibinfo {author} {\bibfnamefont {L.}~\bibnamefont {Childress}},
  \bibinfo {author} {\bibfnamefont {H.}~\bibnamefont {Bernien}}, \bibinfo
  {author} {\bibfnamefont {B.}~\bibnamefont {Hensen}}, \bibinfo {author}
  {\bibfnamefont {P.~F.~A.}\ \bibnamefont {Alkemade}}, \ and\ \bibinfo {author}
  {\bibfnamefont {R.}~\bibnamefont {Hanson}},\ }\href@noop {} {\bibfield
  {journal} {\bibinfo  {journal} {Nature}\ }\textbf {\bibinfo {volume} {477}},\
  \bibinfo {pages} {574} (\bibinfo {year} {2011})}\BibitemShut {NoStop}%
\bibitem [{\citenamefont {Wolters}\ \emph {et~al.}(2010)\citenamefont
  {Wolters}, \citenamefont {Schell}, \citenamefont {Kewes}, \citenamefont
  {N\"{u}sse}, \citenamefont {Schoengen}, \citenamefont {D\"{o}scher},
  \citenamefont {Hannappel}, \citenamefont {L\"{o}chel}, \citenamefont
  {Barth},\ and\ \citenamefont {Benson}}]{Wolters2010}%
  \BibitemOpen
  \bibfield  {author} {\bibinfo {author} {\bibfnamefont {J.}~\bibnamefont
  {Wolters}}, \bibinfo {author} {\bibfnamefont {A.~W.}\ \bibnamefont {Schell}},
  \bibinfo {author} {\bibfnamefont {G.}~\bibnamefont {Kewes}}, \bibinfo
  {author} {\bibfnamefont {N.}~\bibnamefont {N\"{u}sse}}, \bibinfo {author}
  {\bibfnamefont {M.}~\bibnamefont {Schoengen}}, \bibinfo {author}
  {\bibfnamefont {H.}~\bibnamefont {D\"{o}scher}}, \bibinfo {author}
  {\bibfnamefont {T.}~\bibnamefont {Hannappel}}, \bibinfo {author}
  {\bibfnamefont {B.}~\bibnamefont {L\"{o}chel}}, \bibinfo {author}
  {\bibfnamefont {M.}~\bibnamefont {Barth}}, \ and\ \bibinfo {author}
  {\bibfnamefont {O.}~\bibnamefont {Benson}},\ }\href {\doibase
  http://dx.doi.org/10.1063/1.3499300} {\bibfield  {journal} {\bibinfo
  {journal} {Applied Physics Letters}\ }\textbf {\bibinfo {volume} {97}},\
  \bibinfo {pages} {141108} (\bibinfo {year} {2010})}\BibitemShut {NoStop}%
\bibitem [{\citenamefont {Beha}\ \emph {et~al.}(2012)\citenamefont {Beha},
  \citenamefont {Fedder}, \citenamefont {Wolfer}, \citenamefont {Becker},
  \citenamefont {Siyushev}, \citenamefont {Jamali}, \citenamefont {Batalov},
  \citenamefont {Hinz}, \citenamefont {Hees}, \citenamefont {Kirste},
  \citenamefont {Obloh}, \citenamefont {Gheeraert}, \citenamefont {Naydenov},
  \citenamefont {Jakobi}, \citenamefont {Dolde}, \citenamefont {Pezzagna},
  \citenamefont {Twittchen}, \citenamefont {Markham}, \citenamefont {Dregely},
  \citenamefont {Giessen}, \citenamefont {Meijer}, \citenamefont {Jelezko},
  \citenamefont {Nebel}, \citenamefont {Bratschitsch}, \citenamefont
  {Leitenstorfer},\ and\ \citenamefont {Wrachtrup}}]{Wrachtrup_12}%
  \BibitemOpen
  \bibfield  {author} {\bibinfo {author} {\bibfnamefont {K.}~\bibnamefont
  {Beha}}, \bibinfo {author} {\bibfnamefont {H.}~\bibnamefont {Fedder}},
  \bibinfo {author} {\bibfnamefont {M.}~\bibnamefont {Wolfer}}, \bibinfo
  {author} {\bibfnamefont {M.~C.}\ \bibnamefont {Becker}}, \bibinfo {author}
  {\bibfnamefont {P.}~\bibnamefont {Siyushev}}, \bibinfo {author}
  {\bibfnamefont {M.}~\bibnamefont {Jamali}}, \bibinfo {author} {\bibfnamefont
  {A.}~\bibnamefont {Batalov}}, \bibinfo {author} {\bibfnamefont
  {C.}~\bibnamefont {Hinz}}, \bibinfo {author} {\bibfnamefont {J.}~\bibnamefont
  {Hees}}, \bibinfo {author} {\bibfnamefont {L.}~\bibnamefont {Kirste}},
  \bibinfo {author} {\bibfnamefont {H.}~\bibnamefont {Obloh}}, \bibinfo
  {author} {\bibfnamefont {E.}~\bibnamefont {Gheeraert}}, \bibinfo {author}
  {\bibfnamefont {B.}~\bibnamefont {Naydenov}}, \bibinfo {author}
  {\bibfnamefont {I.}~\bibnamefont {Jakobi}}, \bibinfo {author} {\bibfnamefont
  {F.}~\bibnamefont {Dolde}}, \bibinfo {author} {\bibfnamefont
  {S.}~\bibnamefont {Pezzagna}}, \bibinfo {author} {\bibfnamefont
  {D.}~\bibnamefont {Twittchen}}, \bibinfo {author} {\bibfnamefont
  {M.}~\bibnamefont {Markham}}, \bibinfo {author} {\bibfnamefont
  {D.}~\bibnamefont {Dregely}}, \bibinfo {author} {\bibfnamefont
  {H.}~\bibnamefont {Giessen}}, \bibinfo {author} {\bibfnamefont
  {J.}~\bibnamefont {Meijer}}, \bibinfo {author} {\bibfnamefont
  {F.}~\bibnamefont {Jelezko}}, \bibinfo {author} {\bibfnamefont {C.~E.}\
  \bibnamefont {Nebel}}, \bibinfo {author} {\bibfnamefont {R.}~\bibnamefont
  {Bratschitsch}}, \bibinfo {author} {\bibfnamefont {A.}~\bibnamefont
  {Leitenstorfer}}, \ and\ \bibinfo {author} {\bibfnamefont {J.}~\bibnamefont
  {Wrachtrup}},\ }\href@noop {} {\bibfield  {journal} {\bibinfo  {journal}
  {Beilstein J. Nanotechnol.}\ }\textbf {\bibinfo {volume} {3}},\ \bibinfo
  {pages} {895} (\bibinfo {year} {2012})}\BibitemShut {NoStop}%
\bibitem [{\citenamefont {Sauvan}\ \emph {et~al.}(2013)\citenamefont {Sauvan},
  \citenamefont {Hugonin}, \citenamefont {Maksymov},\ and\ \citenamefont
  {Lalanne}}]{Maksymov_PRL_2013}%
  \BibitemOpen
  \bibfield  {author} {\bibinfo {author} {\bibfnamefont {C.}~\bibnamefont
  {Sauvan}}, \bibinfo {author} {\bibfnamefont {J.~P.}\ \bibnamefont {Hugonin}},
  \bibinfo {author} {\bibfnamefont {I.~S.}\ \bibnamefont {Maksymov}}, \ and\
  \bibinfo {author} {\bibfnamefont {P.}~\bibnamefont {Lalanne}},\ }\href@noop
  {} {\bibfield  {journal} {\bibinfo  {journal} {Physical Review Letters}\
  }\textbf {\bibinfo {volume} {110}},\ \bibinfo {pages} {237401} (\bibinfo
  {year} {2013})}\BibitemShut {NoStop}%
\bibitem [{\citenamefont {Biagioni}\ \emph {et~al.}(2012)\citenamefont
  {Biagioni}, \citenamefont {Huang},\ and\ \citenamefont {Hecht}}]{26}%
  \BibitemOpen
  \bibfield  {author} {\bibinfo {author} {\bibfnamefont {P.}~\bibnamefont
  {Biagioni}}, \bibinfo {author} {\bibfnamefont {J.}~\bibnamefont {Huang}}, \
  and\ \bibinfo {author} {\bibfnamefont {B.}~\bibnamefont {Hecht}},\
  }\href@noop {} {\bibfield  {journal} {\bibinfo  {journal} {Rep. Prog. Phys.}\
  }\textbf {\bibinfo {volume} {75}},\ \bibinfo {pages} {024402} (\bibinfo
  {year} {2012})}\BibitemShut {NoStop}%
\bibitem [{\citenamefont {Purcell}(1946)}]{Purcell_46}%
  \BibitemOpen
  \bibfield  {author} {\bibinfo {author} {\bibfnamefont {E.~M.}\ \bibnamefont
  {Purcell}},\ }\href@noop {} {\bibfield  {journal} {\bibinfo  {journal} {Phys.
  Rev.}\ }\textbf {\bibinfo {volume} {69}},\ \bibinfo {pages} {681} (\bibinfo
  {year} {1946})}\BibitemShut {NoStop}%
\bibitem [{\citenamefont {Kinkhabwala}\ \emph {et~al.}(2009)\citenamefont
  {Kinkhabwala}, \citenamefont {Yu}, \citenamefont {Fan}, \citenamefont
  {Avlasevich}, \citenamefont {Mullen},\ and\ \citenamefont
  {Moerner}}]{Moerner_bowtie_09}%
  \BibitemOpen
  \bibfield  {author} {\bibinfo {author} {\bibfnamefont {A.}~\bibnamefont
  {Kinkhabwala}}, \bibinfo {author} {\bibfnamefont {Z.}~\bibnamefont {Yu}},
  \bibinfo {author} {\bibfnamefont {S.}~\bibnamefont {Fan}}, \bibinfo {author}
  {\bibfnamefont {Y.}~\bibnamefont {Avlasevich}}, \bibinfo {author}
  {\bibfnamefont {K.}~\bibnamefont {Mullen}}, \ and\ \bibinfo {author}
  {\bibfnamefont {W.~E.}\ \bibnamefont {Moerner}},\ }\href@noop {} {\bibfield
  {journal} {\bibinfo  {journal} {Nat. Phot.}\ }\textbf {\bibinfo {volume}
  {3}},\ \bibinfo {pages} {654} (\bibinfo {year} {2009})}\BibitemShut {NoStop}%
\bibitem [{\citenamefont {Vahala}(2003)}]{Vahala_2003}%
  \BibitemOpen
  \bibfield  {author} {\bibinfo {author} {\bibfnamefont {K.~J.}\ \bibnamefont
  {Vahala}},\ }\href@noop {} {\bibfield  {journal} {\bibinfo  {journal}
  {Nature}\ }\textbf {\bibinfo {volume} {424}},\ \bibinfo {pages} {839}
  (\bibinfo {year} {2003})}\BibitemShut {NoStop}%
\bibitem [{\citenamefont {Krasnok}\ \emph {et~al.}(2013)\citenamefont
  {Krasnok}, \citenamefont {Maksymov}, \citenamefont {Denisyuk}, \citenamefont
  {Belov}, \citenamefont {Miroshnichenko}, \citenamefont {Simovski},\ and\
  \citenamefont {Kivshar}}]{58}%
  \BibitemOpen
  \bibfield  {author} {\bibinfo {author} {\bibfnamefont {A.~E.}\ \bibnamefont
  {Krasnok}}, \bibinfo {author} {\bibfnamefont {I.~S.}\ \bibnamefont
  {Maksymov}}, \bibinfo {author} {\bibfnamefont {A.~I.}\ \bibnamefont
  {Denisyuk}}, \bibinfo {author} {\bibfnamefont {P.~A.}\ \bibnamefont {Belov}},
  \bibinfo {author} {\bibfnamefont {A.~E.}\ \bibnamefont {Miroshnichenko}},
  \bibinfo {author} {\bibfnamefont {C.~R.}\ \bibnamefont {Simovski}}, \ and\
  \bibinfo {author} {\bibfnamefont {Y.~S.}\ \bibnamefont {Kivshar}},\
  }\href@noop {} {\bibfield  {journal} {\bibinfo  {journal} {Phys.-Usp.}\
  }\textbf {\bibinfo {volume} {56}},\ \bibinfo {pages} {539} (\bibinfo {year}
  {2013})}\BibitemShut {NoStop}%
\bibitem [{\citenamefont {Englund}\ \emph {et~al.}(2010)\citenamefont
  {Englund}, \citenamefont {Shields}, \citenamefont {Rivoire}, \citenamefont
  {Hatami}, \citenamefont {Vuckovic}, \citenamefont {Park},\ and\ \citenamefont
  {Lukin}}]{Englund2010}%
  \BibitemOpen
  \bibfield  {author} {\bibinfo {author} {\bibfnamefont {D.}~\bibnamefont
  {Englund}}, \bibinfo {author} {\bibfnamefont {B.}~\bibnamefont {Shields}},
  \bibinfo {author} {\bibfnamefont {K.}~\bibnamefont {Rivoire}}, \bibinfo
  {author} {\bibfnamefont {F.}~\bibnamefont {Hatami}}, \bibinfo {author}
  {\bibfnamefont {J.}~\bibnamefont {Vuckovic}}, \bibinfo {author}
  {\bibfnamefont {H.}~\bibnamefont {Park}}, \ and\ \bibinfo {author}
  {\bibfnamefont {M.~D.}\ \bibnamefont {Lukin}},\ }\href@noop {} {\bibfield
  {journal} {\bibinfo  {journal} {Nano Letters}\ }\textbf {\bibinfo {volume}
  {10}},\ \bibinfo {pages} {3922} (\bibinfo {year} {2010})}\BibitemShut
  {NoStop}%
\bibitem [{\citenamefont {Faraon}\ \emph {et~al.}(2012)\citenamefont {Faraon},
  \citenamefont {Santori}, \citenamefont {Huang}, \citenamefont {Acosta},\ and\
  \citenamefont {Beausoleil}}]{Faraon2012}%
  \BibitemOpen
  \bibfield  {author} {\bibinfo {author} {\bibfnamefont {A.}~\bibnamefont
  {Faraon}}, \bibinfo {author} {\bibfnamefont {C.}~\bibnamefont {Santori}},
  \bibinfo {author} {\bibfnamefont {Z.}~\bibnamefont {Huang}}, \bibinfo
  {author} {\bibfnamefont {V.~M.}\ \bibnamefont {Acosta}}, \ and\ \bibinfo
  {author} {\bibfnamefont {R.~G.}\ \bibnamefont {Beausoleil}},\ }\href@noop {}
  {\bibfield  {journal} {\bibinfo  {journal} {Phys. Rev. Lett.}\ }\textbf
  {\bibinfo {volume} {109}},\ \bibinfo {pages} {33604} (\bibinfo {year}
  {2012})},\ \Eprint {http://arxiv.org/abs/arXiv:1202.0806v1}
  {arXiv:arXiv:1202.0806v1} \BibitemShut {NoStop}%
\bibitem [{\citenamefont {Fu}\ \emph {et~al.}(2008)\citenamefont {Fu},
  \citenamefont {Santori}, \citenamefont {Barclay}, \citenamefont
  {Aharonovich}, \citenamefont {Prawer}, \citenamefont {Meyer}, \citenamefont
  {Holm},\ and\ \citenamefont {Beausoleil}}]{Fu2008}%
  \BibitemOpen
  \bibfield  {author} {\bibinfo {author} {\bibfnamefont {K.-M.~C.}\
  \bibnamefont {Fu}}, \bibinfo {author} {\bibfnamefont {C.}~\bibnamefont
  {Santori}}, \bibinfo {author} {\bibfnamefont {P.~E.}\ \bibnamefont
  {Barclay}}, \bibinfo {author} {\bibfnamefont {I.}~\bibnamefont
  {Aharonovich}}, \bibinfo {author} {\bibfnamefont {S.}~\bibnamefont {Prawer}},
  \bibinfo {author} {\bibfnamefont {N.}~\bibnamefont {Meyer}}, \bibinfo
  {author} {\bibfnamefont {A.~M.}\ \bibnamefont {Holm}}, \ and\ \bibinfo
  {author} {\bibfnamefont {R.~G.}\ \bibnamefont {Beausoleil}},\ }\href@noop {}
  {\bibfield  {journal} {\bibinfo  {journal} {Appl. Phys. Lett.}\ }\textbf
  {\bibinfo {volume} {93}} (\bibinfo {year} {2008})},\ \Eprint
  {http://arxiv.org/abs/arXiv:0811.0328v1} {arXiv:arXiv:0811.0328v1}
  \BibitemShut {NoStop}%
\bibitem [{\citenamefont {Barclay}\ \emph {et~al.}(2011)\citenamefont
  {Barclay}, \citenamefont {Fu}, \citenamefont {Santori}, \citenamefont
  {Faraon},\ and\ \citenamefont {Beausoleil}}]{Barclay2011}%
  \BibitemOpen
  \bibfield  {author} {\bibinfo {author} {\bibfnamefont {P.~E.}\ \bibnamefont
  {Barclay}}, \bibinfo {author} {\bibfnamefont {K.-M.~C.}\ \bibnamefont {Fu}},
  \bibinfo {author} {\bibfnamefont {C.}~\bibnamefont {Santori}}, \bibinfo
  {author} {\bibfnamefont {A.}~\bibnamefont {Faraon}}, \ and\ \bibinfo {author}
  {\bibfnamefont {R.~G.}\ \bibnamefont {Beausoleil}},\ }\href@noop {}
  {\bibfield  {journal} {\bibinfo  {journal} {Physical Review X}\ }\textbf
  {\bibinfo {volume} {1}},\ \bibinfo {pages} {011007} (\bibinfo {year}
  {2011})}\BibitemShut {NoStop}%
\bibitem [{\citenamefont {Faraon}\ \emph {et~al.}(2011)\citenamefont {Faraon},
  \citenamefont {Barclay}, \citenamefont {Santori}, \citenamefont {Fu},\ and\
  \citenamefont {Beausoleil}}]{Beausoleil_11}%
  \BibitemOpen
  \bibfield  {author} {\bibinfo {author} {\bibfnamefont {A.}~\bibnamefont
  {Faraon}}, \bibinfo {author} {\bibfnamefont {P.~E.}\ \bibnamefont {Barclay}},
  \bibinfo {author} {\bibfnamefont {C.}~\bibnamefont {Santori}}, \bibinfo
  {author} {\bibfnamefont {K.-M.~C.}\ \bibnamefont {Fu}}, \ and\ \bibinfo
  {author} {\bibfnamefont {R.~G.}\ \bibnamefont {Beausoleil}},\ }\href@noop {}
  {\bibfield  {journal} {\bibinfo  {journal} {Nature Photonics}\ }\textbf
  {\bibinfo {volume} {5}},\ \bibinfo {pages} {301} (\bibinfo {year}
  {2011})}\BibitemShut {NoStop}%
\bibitem [{\citenamefont {Chi}\ \emph {et~al.}(2011)\citenamefont {Chi},
  \citenamefont {Chen}, \citenamefont {Jelezko},\ and\ \citenamefont
  {Zeng}}]{Chi2011}%
  \BibitemOpen
  \bibfield  {author} {\bibinfo {author} {\bibfnamefont {Y.}~\bibnamefont
  {Chi}}, \bibinfo {author} {\bibfnamefont {G.}~\bibnamefont {Chen}}, \bibinfo
  {author} {\bibfnamefont {F.}~\bibnamefont {Jelezko}}, \ and\ \bibinfo
  {author} {\bibfnamefont {H.}~\bibnamefont {Zeng}},\ }\href@noop {} {\bibfield
   {journal} {\bibinfo  {journal} {IEEE Photonics Technology Letters}\ }\textbf
  {\bibinfo {volume} {23}},\ \bibinfo {pages} {374} (\bibinfo {year}
  {2011})}\BibitemShut {NoStop}%
\bibitem [{\citenamefont {Hausmann}\ \emph {et~al.}(2010)\citenamefont
  {Hausmann}, \citenamefont {Khan}, \citenamefont {Zhang}, \citenamefont
  {Babinec}, \citenamefont {Martinick}, \citenamefont {McCutcheon},
  \citenamefont {Hemmer},\ and\ \citenamefont {Loncar}}]{Hausmann2010}%
  \BibitemOpen
  \bibfield  {author} {\bibinfo {author} {\bibfnamefont {B.~J.}\ \bibnamefont
  {Hausmann}}, \bibinfo {author} {\bibfnamefont {M.}~\bibnamefont {Khan}},
  \bibinfo {author} {\bibfnamefont {Y.}~\bibnamefont {Zhang}}, \bibinfo
  {author} {\bibfnamefont {T.~M.}\ \bibnamefont {Babinec}}, \bibinfo {author}
  {\bibfnamefont {K.}~\bibnamefont {Martinick}}, \bibinfo {author}
  {\bibfnamefont {M.}~\bibnamefont {McCutcheon}}, \bibinfo {author}
  {\bibfnamefont {P.~R.}\ \bibnamefont {Hemmer}}, \ and\ \bibinfo {author}
  {\bibfnamefont {M.}~\bibnamefont {Loncar}},\ }\href@noop {} {\bibfield
  {journal} {\bibinfo  {journal} {Diamond and Related Materials}\ }\textbf
  {\bibinfo {volume} {19}},\ \bibinfo {pages} {621} (\bibinfo {year}
  {2010})}\BibitemShut {NoStop}%
\bibitem [{\citenamefont {Choy}\ \emph {et~al.}(2011)\citenamefont {Choy},
  \citenamefont {Hausmann},\ and\ \citenamefont {Babinec}}]{Choy2011}%
  \BibitemOpen
  \bibfield  {author} {\bibinfo {author} {\bibfnamefont {J.~T.}\ \bibnamefont
  {Choy}}, \bibinfo {author} {\bibfnamefont {B.~J.~M.}\ \bibnamefont
  {Hausmann}}, \ and\ \bibinfo {author} {\bibfnamefont {T.~M.}\ \bibnamefont
  {Babinec}},\ }\href@noop {} {\bibfield  {journal} {\bibinfo  {journal}
  {Nature Photonics}\ }\textbf {\bibinfo {volume} {5}},\ \bibinfo {pages} {1}
  (\bibinfo {year} {2011})}\BibitemShut {NoStop}%
\bibitem [{\citenamefont {Bulu}\ \emph {et~al.}(2011)\citenamefont {Bulu},
  \citenamefont {Babinec},\ and\ \citenamefont {Hausmann}}]{Bulu2011}%
  \BibitemOpen
  \bibfield  {author} {\bibinfo {author} {\bibfnamefont {I.}~\bibnamefont
  {Bulu}}, \bibinfo {author} {\bibfnamefont {T.}~\bibnamefont {Babinec}}, \
  and\ \bibinfo {author} {\bibfnamefont {B.}~\bibnamefont {Hausmann}},\
  }\href@noop {} {\bibfield  {journal} {\bibinfo  {journal} {Optics Express}\
  }\textbf {\bibinfo {volume} {199}},\ \bibinfo {pages} {1694} (\bibinfo {year}
  {2011})},\ \Eprint {http://arxiv.org/abs/arXiv:1006.2093v2}
  {arXiv:arXiv:1006.2093v2} \BibitemShut {NoStop}%
\bibitem [{\citenamefont {Sage}\ \emph {et~al.}(2012)\citenamefont {Sage},
  \citenamefont {Pham}, \citenamefont {Bar-Gill}, \citenamefont {Belthangady},
  \citenamefont {Lukin}, \citenamefont {Yacoby},\ and\ \citenamefont
  {Walsworth}}]{Sage2012}%
  \BibitemOpen
  \bibfield  {author} {\bibinfo {author} {\bibfnamefont {D.~L.}\ \bibnamefont
  {Sage}}, \bibinfo {author} {\bibfnamefont {L.~M.}\ \bibnamefont {Pham}},
  \bibinfo {author} {\bibfnamefont {N.}~\bibnamefont {Bar-Gill}}, \bibinfo
  {author} {\bibfnamefont {C.}~\bibnamefont {Belthangady}}, \bibinfo {author}
  {\bibfnamefont {M.~D.}\ \bibnamefont {Lukin}}, \bibinfo {author}
  {\bibfnamefont {A.}~\bibnamefont {Yacoby}}, \ and\ \bibinfo {author}
  {\bibfnamefont {R.~L.}\ \bibnamefont {Walsworth}},\ }\href@noop {} {\bibfield
   {journal} {\bibinfo  {journal} {Physical Review B}\ }\textbf {\bibinfo
  {volume} {85}},\ \bibinfo {pages} {121202(R)} (\bibinfo {year} {2012})},\
  \Eprint {http://arxiv.org/abs/arXiv:1201.0674v1} {arXiv:arXiv:1201.0674v1}
  \BibitemShut {NoStop}%
\bibitem [{\citenamefont {Novotny}\ and\ \citenamefont {van
  Hulst}(2011)}]{Novotny_10_NatPhot}%
  \BibitemOpen
  \bibfield  {author} {\bibinfo {author} {\bibfnamefont {L.}~\bibnamefont
  {Novotny}}\ and\ \bibinfo {author} {\bibfnamefont {N.}~\bibnamefont {van
  Hulst}},\ }\href@noop {} {\bibfield  {journal} {\bibinfo  {journal} {Nat.
  Photon.}\ }\textbf {\bibinfo {volume} {5}},\ \bibinfo {pages} {83} (\bibinfo
  {year} {2011})}\BibitemShut {NoStop}%
\bibitem [{\citenamefont {Gramotnev}\ and\ \citenamefont
  {Bozhevolnyi}(2010)}]{Bozhevolnyi_NP_10}%
  \BibitemOpen
  \bibfield  {author} {\bibinfo {author} {\bibfnamefont {D.~K.}\ \bibnamefont
  {Gramotnev}}\ and\ \bibinfo {author} {\bibfnamefont {S.~I.}\ \bibnamefont
  {Bozhevolnyi}},\ }\href@noop {} {\bibfield  {journal} {\bibinfo  {journal}
  {Nature Photonics}\ }\textbf {\bibinfo {volume} {4}},\ \bibinfo {pages} {83}
  (\bibinfo {year} {2010})}\BibitemShut {NoStop}%
\bibitem [{\citenamefont {Hess}\ \emph {et~al.}(2012)\citenamefont {Hess},
  \citenamefont {Pendry}, \citenamefont {Maier}, \citenamefont {Oulton},
  \citenamefont {Hamm},\ and\ \citenamefont {Tsakmakidis}}]{Tsakmakidis_NM_12}%
  \BibitemOpen
  \bibfield  {author} {\bibinfo {author} {\bibfnamefont {O.}~\bibnamefont
  {Hess}}, \bibinfo {author} {\bibfnamefont {J.~B.}\ \bibnamefont {Pendry}},
  \bibinfo {author} {\bibfnamefont {S.~A.}\ \bibnamefont {Maier}}, \bibinfo
  {author} {\bibfnamefont {R.~F.}\ \bibnamefont {Oulton}}, \bibinfo {author}
  {\bibfnamefont {J.~M.}\ \bibnamefont {Hamm}}, \ and\ \bibinfo {author}
  {\bibfnamefont {K.~L.}\ \bibnamefont {Tsakmakidis}},\ }\href@noop {}
  {\bibfield  {journal} {\bibinfo  {journal} {Nature Materials}\ }\textbf
  {\bibinfo {volume} {11}},\ \bibinfo {pages} {573} (\bibinfo {year}
  {2012})}\BibitemShut {NoStop}%
\bibitem [{\citenamefont {Gramotnev}\ and\ \citenamefont
  {Bozhevolnyi}(2014)}]{Bozhevolnyi_NP_14}%
  \BibitemOpen
  \bibfield  {author} {\bibinfo {author} {\bibfnamefont {D.~K.}\ \bibnamefont
  {Gramotnev}}\ and\ \bibinfo {author} {\bibfnamefont {S.~I.}\ \bibnamefont
  {Bozhevolnyi}},\ }\href@noop {} {\bibfield  {journal} {\bibinfo  {journal}
  {Nature Photonics}\ }\textbf {\bibinfo {volume} {8}},\ \bibinfo {pages} {13}
  (\bibinfo {year} {2014})}\BibitemShut {NoStop}%
\bibitem [{\citenamefont {Krasnok}\ \emph {et~al.}(2012)\citenamefont
  {Krasnok}, \citenamefont {Miroshnichenko}, \citenamefont {Belov},\ and\
  \citenamefont {Kivshar}}]{7}%
  \BibitemOpen
  \bibfield  {author} {\bibinfo {author} {\bibfnamefont {A.~E.}\ \bibnamefont
  {Krasnok}}, \bibinfo {author} {\bibfnamefont {A.~E.}\ \bibnamefont
  {Miroshnichenko}}, \bibinfo {author} {\bibfnamefont {P.~A.}\ \bibnamefont
  {Belov}}, \ and\ \bibinfo {author} {\bibfnamefont {Y.~S.}\ \bibnamefont
  {Kivshar}},\ }\href@noop {} {\bibfield  {journal} {\bibinfo  {journal} {Opt.
  Express}\ }\textbf {\bibinfo {volume} {20}},\ \bibinfo {pages} {20599}
  (\bibinfo {year} {2012})}\BibitemShut {NoStop}%
\bibitem [{\citenamefont {Rolly}\ \emph {et~al.}(2012)\citenamefont {Rolly},
  \citenamefont {Stout},\ and\ \citenamefont {Bonod}}]{BonodOE}%
  \BibitemOpen
  \bibfield  {author} {\bibinfo {author} {\bibfnamefont {B.}~\bibnamefont
  {Rolly}}, \bibinfo {author} {\bibfnamefont {B.}~\bibnamefont {Stout}}, \ and\
  \bibinfo {author} {\bibfnamefont {N.}~\bibnamefont {Bonod}},\ }\href@noop {}
  {\bibfield  {journal} {\bibinfo  {journal} {Optics Exp.}\ }\textbf {\bibinfo
  {volume} {20}},\ \bibinfo {pages} {20376} (\bibinfo {year}
  {2012})}\BibitemShut {NoStop}%
\bibitem [{\citenamefont {Rolly}\ \emph {et~al.}(2013)\citenamefont {Rolly},
  \citenamefont {Geffrin}, \citenamefont {Abdeddaim}, \citenamefont {Stout},\
  and\ \citenamefont {Bonod}}]{BonodScRep}%
  \BibitemOpen
  \bibfield  {author} {\bibinfo {author} {\bibfnamefont {B.}~\bibnamefont
  {Rolly}}, \bibinfo {author} {\bibfnamefont {J.-M.}\ \bibnamefont {Geffrin}},
  \bibinfo {author} {\bibfnamefont {R.}~\bibnamefont {Abdeddaim}}, \bibinfo
  {author} {\bibfnamefont {B.}~\bibnamefont {Stout}}, \ and\ \bibinfo {author}
  {\bibfnamefont {N.}~\bibnamefont {Bonod}},\ }\href@noop {} {\bibfield
  {journal} {\bibinfo  {journal} {Scientific Reports}\ }\textbf {\bibinfo
  {volume} {3}},\ \bibinfo {pages} {3063} (\bibinfo {year} {2013})}\BibitemShut
  {NoStop}%
\bibitem [{\citenamefont {Evlyukhin}\ \emph {et~al.}(2014)\citenamefont
  {Evlyukhin}, \citenamefont {Eriksen}, \citenamefont {Cheng}, \citenamefont
  {Beermann}, \citenamefont {Reinhardt}, \citenamefont {Petrov}, \citenamefont
  {Prorok}, \citenamefont {Eich}, \citenamefont {Chichkov},\ and\ \citenamefont
  {Bozhevolnyi}}]{Bozhevolnyi_SR_14}%
  \BibitemOpen
  \bibfield  {author} {\bibinfo {author} {\bibfnamefont {A.~B.}\ \bibnamefont
  {Evlyukhin}}, \bibinfo {author} {\bibfnamefont {R.~L.}\ \bibnamefont
  {Eriksen}}, \bibinfo {author} {\bibfnamefont {W.}~\bibnamefont {Cheng}},
  \bibinfo {author} {\bibfnamefont {J.}~\bibnamefont {Beermann}}, \bibinfo
  {author} {\bibfnamefont {C.}~\bibnamefont {Reinhardt}}, \bibinfo {author}
  {\bibfnamefont {A.}~\bibnamefont {Petrov}}, \bibinfo {author} {\bibfnamefont
  {S.}~\bibnamefont {Prorok}}, \bibinfo {author} {\bibfnamefont
  {M.}~\bibnamefont {Eich}}, \bibinfo {author} {\bibfnamefont {B.~N.}\
  \bibnamefont {Chichkov}}, \ and\ \bibinfo {author} {\bibfnamefont {S.~I.}\
  \bibnamefont {Bozhevolnyi}},\ }\href@noop {} {\bibfield  {journal} {\bibinfo
  {journal} {Scientific Reports}\ }\textbf {\bibinfo {volume} {4}},\ \bibinfo
  {pages} {4126} (\bibinfo {year} {2014})}\BibitemShut {NoStop}%
\bibitem [{\citenamefont {Kuznetsov}\ \emph {et~al.}(2012)\citenamefont
  {Kuznetsov}, \citenamefont {Miroshnichenko}, \citenamefont {Fu},
  \citenamefont {Zhang},\ and\ \citenamefont {Lukyanchuk}}]{Kuznetsov}%
  \BibitemOpen
  \bibfield  {author} {\bibinfo {author} {\bibfnamefont {A.~I.}\ \bibnamefont
  {Kuznetsov}}, \bibinfo {author} {\bibfnamefont {A.~E.}\ \bibnamefont
  {Miroshnichenko}}, \bibinfo {author} {\bibfnamefont {Y.~H.}\ \bibnamefont
  {Fu}}, \bibinfo {author} {\bibfnamefont {J.}~\bibnamefont {Zhang}}, \ and\
  \bibinfo {author} {\bibfnamefont {B.}~\bibnamefont {Lukyanchuk}},\
  }\href@noop {} {\bibfield  {journal} {\bibinfo  {journal} {Sci. Rep.}\
  }\textbf {\bibinfo {volume} {2}},\ \bibinfo {pages} {492} (\bibinfo {year}
  {2012})}\BibitemShut {NoStop}%
\bibitem [{\citenamefont {Evlyukhin}\ \emph {et~al.}(2012)\citenamefont
  {Evlyukhin}, \citenamefont {Novikov}, \citenamefont {Zywietz}, \citenamefont
  {Eriksen}, \citenamefont {Reinhardt}, \citenamefont {Bozhevolnyi},\ and\
  \citenamefont {Chichkov}}]{Chichkov_NL_12}%
  \BibitemOpen
  \bibfield  {author} {\bibinfo {author} {\bibfnamefont {A.~B.}\ \bibnamefont
  {Evlyukhin}}, \bibinfo {author} {\bibfnamefont {S.~M.}\ \bibnamefont
  {Novikov}}, \bibinfo {author} {\bibfnamefont {U.}~\bibnamefont {Zywietz}},
  \bibinfo {author} {\bibfnamefont {R.~L.}\ \bibnamefont {Eriksen}}, \bibinfo
  {author} {\bibfnamefont {C.}~\bibnamefont {Reinhardt}}, \bibinfo {author}
  {\bibfnamefont {S.~I.}\ \bibnamefont {Bozhevolnyi}}, \ and\ \bibinfo {author}
  {\bibfnamefont {B.~N.}\ \bibnamefont {Chichkov}},\ }\href@noop {} {\bibfield
  {journal} {\bibinfo  {journal} {Nano Lett.}\ }\textbf {\bibinfo {volume}
  {12}},\ \bibinfo {pages} {3749} (\bibinfo {year} {2012})}\BibitemShut
  {NoStop}%
\bibitem [{\citenamefont {Evlyukhin}\ \emph {et~al.}(2010)\citenamefont
  {Evlyukhin}, \citenamefont {Reinhardt}, \citenamefont {Seidel}, \citenamefont
  {Luk’yanchuk},\ and\ \citenamefont {Chichkov}}]{Chichkov_PRB_10}%
  \BibitemOpen
  \bibfield  {author} {\bibinfo {author} {\bibfnamefont {A.~B.}\ \bibnamefont
  {Evlyukhin}}, \bibinfo {author} {\bibfnamefont {C.}~\bibnamefont
  {Reinhardt}}, \bibinfo {author} {\bibfnamefont {A.}~\bibnamefont {Seidel}},
  \bibinfo {author} {\bibfnamefont {B.~S.}\ \bibnamefont {Luk’yanchuk}}, \ and\
  \bibinfo {author} {\bibfnamefont {B.~N.}\ \bibnamefont {Chichkov}},\
  }\href@noop {} {\bibfield  {journal} {\bibinfo  {journal} {Phys. Rev. B}\
  }\textbf {\bibinfo {volume} {82}},\ \bibinfo {pages} {045404} (\bibinfo
  {year} {2010})}\BibitemShut {NoStop}%
\bibitem [{\citenamefont {Krasnok}\ \emph {et~al.}(2014)\citenamefont
  {Krasnok}, \citenamefont {Simovski}, \citenamefont {Belov},\ and\
  \citenamefont {Kivshar}}]{KrasnokNanoscale}%
  \BibitemOpen
  \bibfield  {author} {\bibinfo {author} {\bibfnamefont {A.~E.}\ \bibnamefont
  {Krasnok}}, \bibinfo {author} {\bibfnamefont {C.~R.}\ \bibnamefont
  {Simovski}}, \bibinfo {author} {\bibfnamefont {P.~A.}\ \bibnamefont {Belov}},
  \ and\ \bibinfo {author} {\bibfnamefont {Y.~S.}\ \bibnamefont {Kivshar}},\
  }\href@noop {} {\bibfield  {journal} {\bibinfo  {journal} {Nanoscale}\
  }\textbf {\bibinfo {volume} {6}},\ \bibinfo {pages} {7354} (\bibinfo {year}
  {2014})}\BibitemShut {NoStop}%
\bibitem [{\citenamefont {Palik}(1985)}]{Palik}%
  \BibitemOpen
  \bibfield  {author} {\bibinfo {author} {\bibfnamefont {E.}~\bibnamefont
  {Palik}},\ }\href@noop {} {\emph {\bibinfo {title} {Handbook of Optical
  Constant of Solids}}}\ (\bibinfo  {publisher} {San Diego, Academic},\
  \bibinfo {year} {1985})\BibitemShut {NoStop}%
\bibitem [{\citenamefont {Sacanna}\ and\ \citenamefont
  {Pine}(2011)}]{Sacanna2011}%
  \BibitemOpen
  \bibfield  {author} {\bibinfo {author} {\bibfnamefont {S.}~\bibnamefont
  {Sacanna}}\ and\ \bibinfo {author} {\bibfnamefont {D.~J.}\ \bibnamefont
  {Pine}},\ }\href@noop {} {\bibfield  {journal} {\bibinfo  {journal} {Current
  Opinion in Colloid and Interface Science}\ }\textbf {\bibinfo {volume}
  {16}},\ \bibinfo {pages} {96} (\bibinfo {year} {2011})}\BibitemShut {NoStop}%
\bibitem [{\citenamefont {Yang}\ \emph {et~al.}(2008)\citenamefont {Yang},
  \citenamefont {Kim}, \citenamefont {Lima},\ and\ \citenamefont
  {Yi}}]{Yang08}%
  \BibitemOpen
  \bibfield  {author} {\bibinfo {author} {\bibfnamefont {S.-M.}\ \bibnamefont
  {Yang}}, \bibinfo {author} {\bibfnamefont {S.-H.}\ \bibnamefont {Kim}},
  \bibinfo {author} {\bibfnamefont {J.-M.}\ \bibnamefont {Lima}}, \ and\
  \bibinfo {author} {\bibfnamefont {G.-R.}\ \bibnamefont {Yi}},\ }\href@noop {}
  {\bibfield  {journal} {\bibinfo  {journal} {J. Mater. Chem.}\ }\textbf
  {\bibinfo {volume} {18}},\ \bibinfo {pages} {2177} (\bibinfo {year}
  {2008})}\BibitemShut {NoStop}%
\bibitem [{\citenamefont {Sacanna}\ \emph {et~al.}(2011)\citenamefont
  {Sacanna}, \citenamefont {Irvine}, \citenamefont {Rossib},\ and\
  \citenamefont {Pinea}}]{Sacanna2011a}%
  \BibitemOpen
  \bibfield  {author} {\bibinfo {author} {\bibfnamefont {S.}~\bibnamefont
  {Sacanna}}, \bibinfo {author} {\bibfnamefont {W.~T.~M.}\ \bibnamefont
  {Irvine}}, \bibinfo {author} {\bibfnamefont {L.}~\bibnamefont {Rossib}}, \
  and\ \bibinfo {author} {\bibfnamefont {D.~J.}\ \bibnamefont {Pinea}},\
  }\href@noop {} {\bibfield  {journal} {\bibinfo  {journal} {Soft Matter}\
  }\textbf {\bibinfo {volume} {7}},\ \bibinfo {pages} {1631} (\bibinfo {year}
  {2011})}\BibitemShut {NoStop}%
\bibitem [{\citenamefont {Dorpe}\ and\ \citenamefont
  {Ye}(2011)}]{VanDorpe2011}%
  \BibitemOpen
  \bibfield  {author} {\bibinfo {author} {\bibfnamefont {P.~V.}\ \bibnamefont
  {Dorpe}}\ and\ \bibinfo {author} {\bibfnamefont {J.}~\bibnamefont {Ye}},\
  }\href@noop {} {\bibfield  {journal} {\bibinfo  {journal} {ACS Nano}\
  }\textbf {\bibinfo {volume} {5}},\ \bibinfo {pages} {6774} (\bibinfo {year}
  {2011})}\BibitemShut {NoStop}%
\bibitem [{\citenamefont {Zhang}\ \emph {et~al.}(2011)\citenamefont {Zhang},
  \citenamefont {Grady}, \citenamefont {Ayala-Orozco},\ and\ \citenamefont
  {Halas}}]{Halas2011}%
  \BibitemOpen
  \bibfield  {author} {\bibinfo {author} {\bibfnamefont {Y.}~\bibnamefont
  {Zhang}}, \bibinfo {author} {\bibfnamefont {N.~K.}\ \bibnamefont {Grady}},
  \bibinfo {author} {\bibfnamefont {C.}~\bibnamefont {Ayala-Orozco}}, \ and\
  \bibinfo {author} {\bibfnamefont {N.~J.}\ \bibnamefont {Halas}},\ }\href@noop
  {} {\bibfield  {journal} {\bibinfo  {journal} {Nano Lett.}\ }\textbf
  {\bibinfo {volume} {11}},\ \bibinfo {pages} {5519} (\bibinfo {year}
  {2011})}\BibitemShut {NoStop}%
\bibitem [{\citenamefont {Wollet}\ \emph {et~al.}(2012)\citenamefont {Wollet},
  \citenamefont {Frank}, \citenamefont {Schaferling}, \citenamefont {Mesch},
  \citenamefont {Hein},\ and\ \citenamefont {Giessen}}]{Giessen2012}%
  \BibitemOpen
  \bibfield  {author} {\bibinfo {author} {\bibfnamefont {L.}~\bibnamefont
  {Wollet}}, \bibinfo {author} {\bibfnamefont {B.}~\bibnamefont {Frank}},
  \bibinfo {author} {\bibfnamefont {M.}~\bibnamefont {Schaferling}}, \bibinfo
  {author} {\bibfnamefont {M.}~\bibnamefont {Mesch}}, \bibinfo {author}
  {\bibfnamefont {S.}~\bibnamefont {Hein}}, \ and\ \bibinfo {author}
  {\bibfnamefont {H.}~\bibnamefont {Giessen}},\ }\href@noop {} {\bibfield
  {journal} {\bibinfo  {journal} {Optic. Mater. Exp.}\ }\textbf {\bibinfo
  {volume} {2}},\ \bibinfo {pages} {1384} (\bibinfo {year} {2012})}\BibitemShut
  {NoStop}%
\bibitem [{\citenamefont {Kuznetsov}\ \emph {et~al.}(2014)\citenamefont
  {Kuznetsov}, \citenamefont {Miroshnichenko}, \citenamefont {Fu},
  \citenamefont {Viswanathan}, \citenamefont {Rahmani}, \citenamefont
  {Valuckas}, \citenamefont {Pan}, \citenamefont {Kivshar}, \citenamefont
  {Pickard},\ and\ \citenamefont {Lukyanchuk}}]{Kuznetsov14}%
  \BibitemOpen
  \bibfield  {author} {\bibinfo {author} {\bibfnamefont {A.~I.}\ \bibnamefont
  {Kuznetsov}}, \bibinfo {author} {\bibfnamefont {A.~E.}\ \bibnamefont
  {Miroshnichenko}}, \bibinfo {author} {\bibfnamefont {Y.~H.}\ \bibnamefont
  {Fu}}, \bibinfo {author} {\bibfnamefont {V.}~\bibnamefont {Viswanathan}},
  \bibinfo {author} {\bibfnamefont {M.}~\bibnamefont {Rahmani}}, \bibinfo
  {author} {\bibfnamefont {V.}~\bibnamefont {Valuckas}}, \bibinfo {author}
  {\bibfnamefont {Z.~Y.}\ \bibnamefont {Pan}}, \bibinfo {author} {\bibfnamefont
  {Y.}~\bibnamefont {Kivshar}}, \bibinfo {author} {\bibfnamefont {D.~S.}\
  \bibnamefont {Pickard}}, \ and\ \bibinfo {author} {\bibfnamefont
  {B.}~\bibnamefont {Lukyanchuk}},\ }\href@noop {} {\bibfield  {journal}
  {\bibinfo  {journal} {Nat. Comm.}\ }\textbf {\bibinfo {volume} {5}},\
  \bibinfo {pages} {3104} (\bibinfo {year} {2014})}\BibitemShut {NoStop}%
\bibitem [{\citenamefont {Zywietz}\ \emph
  {et~al.}(2014{\natexlab{a}})\citenamefont {Zywietz}, \citenamefont
  {Evlyukhin}, \citenamefont {Reinhardt},\ and\ \citenamefont
  {Chichkov}}]{Chichkov_NC_14}%
  \BibitemOpen
  \bibfield  {author} {\bibinfo {author} {\bibfnamefont {U.}~\bibnamefont
  {Zywietz}}, \bibinfo {author} {\bibfnamefont {A.~B.}\ \bibnamefont
  {Evlyukhin}}, \bibinfo {author} {\bibfnamefont {C.}~\bibnamefont
  {Reinhardt}}, \ and\ \bibinfo {author} {\bibfnamefont {B.~N.}\ \bibnamefont
  {Chichkov}},\ }\href@noop {} {\bibfield  {journal} {\bibinfo  {journal}
  {Nature Communications}\ }\textbf {\bibinfo {volume} {5}},\ \bibinfo {pages}
  {3402} (\bibinfo {year} {2014}{\natexlab{a}})}\BibitemShut {NoStop}%
\bibitem [{\citenamefont {Barchanski}\ \emph {et~al.}(2014)\citenamefont
  {Barchanski}, \citenamefont {Evlyukhin}, \citenamefont {Koroleva},
  \citenamefont {Reinhardt}, \citenamefont {Sajti}, \citenamefont {Zywietz},\
  and\ \citenamefont {Chichkov}}]{Chichkov_Book}%
  \BibitemOpen
  \bibfield  {author} {\bibinfo {author} {\bibfnamefont {A.}~\bibnamefont
  {Barchanski}}, \bibinfo {author} {\bibfnamefont {A.~B.}\ \bibnamefont
  {Evlyukhin}}, \bibinfo {author} {\bibfnamefont {A.}~\bibnamefont {Koroleva}},
  \bibinfo {author} {\bibfnamefont {C.}~\bibnamefont {Reinhardt}}, \bibinfo
  {author} {\bibfnamefont {C.~L.}\ \bibnamefont {Sajti}}, \bibinfo {author}
  {\bibfnamefont {U.}~\bibnamefont {Zywietz}}, \ and\ \bibinfo {author}
  {\bibfnamefont {B.~N.}\ \bibnamefont {Chichkov}},\ }\href@noop {} {\emph
  {\bibinfo {title} {Fundamentals of Laser-Assisted Micro- and
  Nanotechnologies}}},\ edited by\ \bibinfo {editor} {\bibfnamefont {V.~P.}\
  \bibnamefont {Veiko}}\ and\ \bibinfo {editor} {\bibfnamefont {V.~I.}\
  \bibnamefont {Konov}}\ (\bibinfo  {publisher} {Springer},\ \bibinfo {year}
  {2014})\BibitemShut {NoStop}%
\bibitem [{\citenamefont {Zywietz}\ \emph
  {et~al.}(2014{\natexlab{b}})\citenamefont {Zywietz}, \citenamefont
  {Reinhardt}, \citenamefont {Evlyukhin}, \citenamefont {Birr},\ and\
  \citenamefont {Chichkov}}]{Chichkov_APA}%
  \BibitemOpen
  \bibfield  {author} {\bibinfo {author} {\bibfnamefont {U.}~\bibnamefont
  {Zywietz}}, \bibinfo {author} {\bibfnamefont {C.}~\bibnamefont {Reinhardt}},
  \bibinfo {author} {\bibfnamefont {A.~B.}\ \bibnamefont {Evlyukhin}}, \bibinfo
  {author} {\bibfnamefont {T.}~\bibnamefont {Birr}}, \ and\ \bibinfo {author}
  {\bibfnamefont {B.~N.}\ \bibnamefont {Chichkov}},\ }\href@noop {} {\bibfield
  {journal} {\bibinfo  {journal} {Appl Phys A}\ }\textbf {\bibinfo {volume}
  {114}},\ \bibinfo {pages} {45} (\bibinfo {year}
  {2014}{\natexlab{b}})}\BibitemShut {NoStop}%
\bibitem [{\citenamefont {Beveratos}\ \emph {et~al.}(2001)\citenamefont
  {Beveratos}, \citenamefont {Brouri}, \citenamefont {Gacoin}, \citenamefont
  {Poizat},\ and\ \citenamefont {Grangier}}]{Beveratos2001}%
  \BibitemOpen
  \bibfield  {author} {\bibinfo {author} {\bibfnamefont {A.}~\bibnamefont
  {Beveratos}}, \bibinfo {author} {\bibfnamefont {R.}~\bibnamefont {Brouri}},
  \bibinfo {author} {\bibfnamefont {T.}~\bibnamefont {Gacoin}}, \bibinfo
  {author} {\bibfnamefont {J.-P.}\ \bibnamefont {Poizat}}, \ and\ \bibinfo
  {author} {\bibfnamefont {P.}~\bibnamefont {Grangier}},\ }\href@noop {}
  {\bibfield  {journal} {\bibinfo  {journal} {Physical Review A}\ }\textbf
  {\bibinfo {volume} {64}} (\bibinfo {year} {2001})}\BibitemShut {NoStop}%
\bibitem [{\citenamefont {Greffet}\ \emph {et~al.}(2011)\citenamefont
  {Greffet}, \citenamefont {Hugonin},\ and\ \citenamefont
  {Besbes}}]{Greffet2011}%
  \BibitemOpen
  \bibfield  {author} {\bibinfo {author} {\bibfnamefont {J.~J.}\ \bibnamefont
  {Greffet}}, \bibinfo {author} {\bibfnamefont {J.~P.}\ \bibnamefont
  {Hugonin}}, \ and\ \bibinfo {author} {\bibfnamefont {M.}~\bibnamefont
  {Besbes}},\ }\href@noop {} {\bibfield  {journal} {\bibinfo  {journal}
  {arxiv.org/abs/1107.0502}\ ,\ \bibinfo {pages} {1}} (\bibinfo {year}
  {2011})},\ \Eprint {http://arxiv.org/abs/arXiv:1107.0502v1}
  {arXiv:arXiv:1107.0502v1} \BibitemShut {NoStop}%
\bibitem [{\citenamefont {Mohtashami}\ and\ \citenamefont
  {Koenderink}(2013)}]{KoenderinkNV13}%
  \BibitemOpen
  \bibfield  {author} {\bibinfo {author} {\bibfnamefont {A.}~\bibnamefont
  {Mohtashami}}\ and\ \bibinfo {author} {\bibfnamefont {A.~F.}\ \bibnamefont
  {Koenderink}},\ }\href@noop {} {\bibfield  {journal} {\bibinfo  {journal}
  {New J. Phys.}\ }\textbf {\bibinfo {volume} {15}},\ \bibinfo {pages} {43017}
  (\bibinfo {year} {2013})}\BibitemShut {NoStop}%
\bibitem [{\citenamefont {Smeltzer}\ \emph {et~al.}(2011)\citenamefont
  {Smeltzer}, \citenamefont {Childress},\ and\ \citenamefont
  {Gali}}]{Smeltzer2011}%
  \BibitemOpen
  \bibfield  {author} {\bibinfo {author} {\bibfnamefont {B.}~\bibnamefont
  {Smeltzer}}, \bibinfo {author} {\bibfnamefont {L.}~\bibnamefont {Childress}},
  \ and\ \bibinfo {author} {\bibfnamefont {A.}~\bibnamefont {Gali}},\
  }\href@noop {} {\bibfield  {journal} {\bibinfo  {journal} {New J. Phys.}\
  }\textbf {\bibinfo {volume} {13}},\ \bibinfo {pages} {025021} (\bibinfo
  {year} {2011})}\BibitemShut {NoStop}%
\bibitem [{\citenamefont {Lai}\ \emph {et~al.}(2010)\citenamefont {Lai},
  \citenamefont {Zheng}, \citenamefont {Treussart},\ and\ \citenamefont
  {Roch}}]{DiepLai2010}%
  \BibitemOpen
  \bibfield  {author} {\bibinfo {author} {\bibfnamefont {N.~D.}\ \bibnamefont
  {Lai}}, \bibinfo {author} {\bibfnamefont {D.}~\bibnamefont {Zheng}}, \bibinfo
  {author} {\bibfnamefont {F.}~\bibnamefont {Treussart}}, \ and\ \bibinfo
  {author} {\bibfnamefont {J.-F.}\ \bibnamefont {Roch}},\ }\href@noop {}
  {\bibfield  {journal} {\bibinfo  {journal} {Adv. Nat. Sci.: Nanosci.
  Nanotechnol.}\ }\textbf {\bibinfo {volume} {1}},\ \bibinfo {pages} {015014}
  (\bibinfo {year} {2010})}\BibitemShut {NoStop}%
\bibitem [{\citenamefont {Geiselmann}\ \emph {et~al.}(2013)\citenamefont
  {Geiselmann}, \citenamefont {Juan}, \citenamefont {Renger}, \citenamefont
  {Say}, \citenamefont {Brown}, \citenamefont {de~Abajo}, \citenamefont
  {Koppens},\ and\ \citenamefont {Quidant}}]{Geiselmann2013}%
  \BibitemOpen
  \bibfield  {author} {\bibinfo {author} {\bibfnamefont {M.}~\bibnamefont
  {Geiselmann}}, \bibinfo {author} {\bibfnamefont {M.~L.}\ \bibnamefont
  {Juan}}, \bibinfo {author} {\bibfnamefont {J.}~\bibnamefont {Renger}},
  \bibinfo {author} {\bibfnamefont {J.~M.}\ \bibnamefont {Say}}, \bibinfo
  {author} {\bibfnamefont {L.~J.}\ \bibnamefont {Brown}}, \bibinfo {author}
  {\bibfnamefont {F.~J.~G.}\ \bibnamefont {de~Abajo}}, \bibinfo {author}
  {\bibfnamefont {F.}~\bibnamefont {Koppens}}, \ and\ \bibinfo {author}
  {\bibfnamefont {R.}~\bibnamefont {Quidant}},\ }\href@noop {} {\bibfield
  {journal} {\bibinfo  {journal} {Nature Nanotechnology}\ }\textbf {\bibinfo
  {volume} {8}},\ \bibinfo {pages} {1} (\bibinfo {year} {2013})}\BibitemShut
  {NoStop}%
\bibitem [{\citenamefont {Dolan}\ \emph {et~al.}(2014)\citenamefont {Dolan},
  \citenamefont {Li}, \citenamefont {Storteboom},\ and\ \citenamefont
  {Gu}}]{Dolan2014}%
  \BibitemOpen
  \bibfield  {author} {\bibinfo {author} {\bibfnamefont {P.~R.}\ \bibnamefont
  {Dolan}}, \bibinfo {author} {\bibfnamefont {X.}~\bibnamefont {Li}}, \bibinfo
  {author} {\bibfnamefont {J.}~\bibnamefont {Storteboom}}, \ and\ \bibinfo
  {author} {\bibfnamefont {M.}~\bibnamefont {Gu}},\ }\href@noop {} {\bibfield
  {journal} {\bibinfo  {journal} {Optics Express}\ }\textbf {\bibinfo {volume}
  {22}},\ \bibinfo {pages} {4379} (\bibinfo {year} {2014})}\BibitemShut
  {NoStop}%
\bibitem [{\citenamefont {Epstein}\ \emph {et~al.}(2005)\citenamefont
  {Epstein}, \citenamefont {Mendoza}, \citenamefont {Kato},\ and\ \citenamefont
  {Awschalom}}]{Epstein2005}%
  \BibitemOpen
  \bibfield  {author} {\bibinfo {author} {\bibfnamefont {R.~J.}\ \bibnamefont
  {Epstein}}, \bibinfo {author} {\bibfnamefont {F.~M.}\ \bibnamefont
  {Mendoza}}, \bibinfo {author} {\bibfnamefont {Y.~K.}\ \bibnamefont {Kato}}, \
  and\ \bibinfo {author} {\bibfnamefont {D.~D.}\ \bibnamefont {Awschalom}},\
  }\href {\doibase 10.1038/nphys141} {\bibfield  {journal} {\bibinfo  {journal}
  {Nature Physics}\ }\textbf {\bibinfo {volume} {1}},\ \bibinfo {pages} {94}
  (\bibinfo {year} {2005})}\BibitemShut {NoStop}%
\bibitem [{\citenamefont {Chen}\ \emph {et~al.}(2013)\citenamefont {Chen},
  \citenamefont {Gaathon}, \citenamefont {Trusheim},\ and\ \citenamefont
  {Englund}}]{Englund_NL_13}%
  \BibitemOpen
  \bibfield  {author} {\bibinfo {author} {\bibfnamefont {E.~H.}\ \bibnamefont
  {Chen}}, \bibinfo {author} {\bibfnamefont {O.}~\bibnamefont {Gaathon}},
  \bibinfo {author} {\bibfnamefont {M.~E.}\ \bibnamefont {Trusheim}}, \ and\
  \bibinfo {author} {\bibfnamefont {D.}~\bibnamefont {Englund}},\ }\href@noop
  {} {\bibfield  {journal} {\bibinfo  {journal} {Nano Lett.}\ }\textbf
  {\bibinfo {volume} {13}},\ \bibinfo {pages} {2073} (\bibinfo {year}
  {2013})}\BibitemShut {NoStop}%
\bibitem [{\citenamefont {Gu}\ \emph {et~al.}(2013)\citenamefont {Gu},
  \citenamefont {Cao}, \citenamefont {Castelletto}, \citenamefont
  {Kouskousis},\ and\ \citenamefont {Li}}]{Xiangping_2013}%
  \BibitemOpen
  \bibfield  {author} {\bibinfo {author} {\bibfnamefont {M.}~\bibnamefont
  {Gu}}, \bibinfo {author} {\bibfnamefont {Y.}~\bibnamefont {Cao}}, \bibinfo
  {author} {\bibfnamefont {S.}~\bibnamefont {Castelletto}}, \bibinfo {author}
  {\bibfnamefont {B.}~\bibnamefont {Kouskousis}}, \ and\ \bibinfo {author}
  {\bibfnamefont {X.}~\bibnamefont {Li}},\ }\href@noop {} {\bibfield  {journal}
  {\bibinfo  {journal} {Opt. Express}\ }\textbf {\bibinfo {volume} {21}},\
  \bibinfo {pages} {17639} (\bibinfo {year} {2013})}\BibitemShut {NoStop}%
\end{thebibliography}
%

\end{document}